\documentclass[twocolumn]{aastex701}

\usepackage{ae,aecompl}
\usepackage{subfigure}
\usepackage{graphicx}	
\usepackage{amsmath}	
\usepackage{amssymb}	
\usepackage{color,xcolor}
\usepackage{float}
\usepackage{multirow}
\usepackage{makecell}
\usepackage{xparse}
\usepackage{ulem}

\newcommand{\s}{\ensuremath{\mbox{~s}}}

\newcommand{\msun}{$M_{\odot}$}
\newcommand{\lsun}{$L_{\odot}$}
\newcommand{\um}{$\mu$m}

\NewDocumentCommand{\fengwei}{ m g }{%
    \IfNoValueTF{#2}
        {\textcolor{teal}{[Fengwei: #1]}}
        {{\color{red}\sout{#1}} \textcolor{teal}{[Fengwei: #2]}}
}

\def   \apjs {{\rm {ApJS}}}
\def   \apjl {{\rm {ApJL}}}

\def   \aap {{\rm {A\&A}}}

\shorttitle{Evolution of starless cores in massive clumps}
\shortauthors{Yang et al.}

\begin{document}

\title{\bf 
Evolution of starless cores in massive clumps seen by the ALMA ASHES and QUARKS surveys}

\author{Dongting Yang}
\affiliation{School of physics and astronomy, Yunnan University, Kunming, 650091, PR China}
\email[]{dongting@mail.ynu.edu.cn}

\author{Hong-li Liu}
\affiliation{School of physics and astronomy, Yunnan University, Kunming, 650091, PR China}
\email[show]{hongliliu2012@gmail.com}

\author{Sheng-li Qin}
\affiliation{School of physics and astronomy, Yunnan University, Kunming, 650091, PR China}
\email[show]{qin@ynu.edu.cn}

\author{Tie Liu}
\affiliation{Shanghai Astronomical Observatory, Chinese Academy of Sciences, 80 Nandan Road, Shanghai 200030, Peoples Republic of China}
\email[show]{liutie@shao.ac.cn}

\author{Wenyu Jiao}
\affiliation{Shanghai Astronomical Observatory, Chinese Academy of Sciences, 80 Nandan Road, Shanghai 200030, Peoples Republic of China}
\email{xxx}

\author{Guido Garay}
\affiliation{Departamento de Astronom\'ia, Universidad de Chile, Casilla 36-D, Santiago, Chile}
\affiliation{Chinese Academy of Sciences South America Center for Astronomy, National Astronomical Observatories, Chinese Academy of Sciences, Beijing, 100101, PR China}
\email{xxx}

\author[0000-0002-7125-7685]{Patricio Sanhueza}
\affiliation{Department of Astronomy, School of Science, The University of Tokyo, 7-3-1 Hongo, Bunkyo, Tokyo 113-0033, Japan}
\email{xxx}

\author[0000-0001-5950-1932]{Fengwei Xu}
\affiliation{Max Planck Institute for Astronomy, Königstuhl 17, D-69117 Heidelberg, Germany}
\email{xxx}

\author{Lei Zhu}
\affiliation{Chinese Academy of Sciences South America Center for Astronomy, National Astronomical Observatories, Chinese Academy of Sciences, Beijing, 100101, PR China}
\email{xxx}

\author[0000-0002-8697-9808]{Sami Dib}
\affiliation{Max Planck Institute for Astronomy, Königstuhl 17, D-69117 Heidelberg, Germany}
\email{xxx}

\author{Xindi Tang}
\affiliation{Xinjiang Astronomical Observatory, Chinese Academy of Sciences,
150 Science 1-Stree, Urumqi, Xinjiang 830011, People's Republic of China}
\email{xxx}

\author{Amelia Stutz}
\affiliation{Departamento de Astronom\'ia, Universidad de Concepci\'on, Av. Esteban Iturra s/n, Distrito Universitario, 160-C, Chile}
\affiliation{Max Planck Institute for Astronomy, Königstuhl 17, D-69117 Heidelberg, Germany}
\email{xxx}

\author[0000-0002-0786-7307]{Xiaofeng Mai}
\affiliation{Shanghai Astronomical Observatory, Chinese Academy of Sciences, 80 Nandan Road, Shanghai 200030, Peoples Republic of China}
\email{xxx}

\author{Siju Zhang}
\affiliation{Departamento de Astronom\'ia, Universidad de Chile, Casilla 36-D, Santiago, Chile}
\affiliation{Chinese Academy of Sciences South America Center for Astronomy, National Astronomical Observatories, Chinese Academy of Sciences, Beijing, 100101, PR China}
\email{xxx}

\author{A.Y. Yang}
\affiliation{National Astronomical Observatories, Chinese Academy of Sciences, Beijing, 100101, PR China} 
\affiliation{Key Laboratory of Radio Astronomy and Technology, Chinese Academy of Sciences, A20 Datun Road, Chaoyang District, Beijing, 100101, PR China}
\email{xxx}

\author{Anandmayee Tej}
\affiliation{Indian Institute of Space Science and Technology, Thiruvananthapuram 695 547, Kerala, India}
\email{tej@iist.ac.in}

\author[0000-0003-1275-5251]{Shanghuo Li}
\affiliation{School of Astronomy and Space Science, Nanjing University, 163 Xianlin Avenue, Nanjing 210023, People’s Republic of China}
\affiliation{Key Laboratory of Modern Astronomy and Astrophysics (Nanjing University), Ministry of Education, Nanjing 210023, People’s Republic of China}
\email{xxx}

\author{Xunchuan Liu}
\affiliation{Shanghai Astronomical Observatory, Chinese Academy of Sciences, 80 Nandan Road, Shanghai 200030, Peoples Republic of China}
\email{xxx}

\author[0000-0002-8586-6721]{Pablo Garc\'ia}
\affiliation{Chinese Academy of Sciences South America Center for Astronomy, National Astronomical Observatories, Chinese Academy of Sciences, Beijing, 100101, PR China}
\affiliation{Instituto de Astronom\'ia, Universidad Cat\'olica del Norte, Av. Angamos 0610, Antofagasta, Chile}
\email{xxx}

\author[0000-0002-5809-4834]{Mika Juvela}
\affiliation{Department of Physics, P.O. box 64, FI- 00014, University of Helsinki, Finland}
\email{xxx}

\author[0000-0002-9875-7436]{James O. Chibueze}
\affiliation{Department of Mathematical Sciences, University of South Africa, Cnr Christian de Wet Rd and Pioneer Avenue, Florida Park, 1709, Roodepoort, South Africa}
\affiliation{Department of Physics and Astronomy, Faculty of Physical Sciences, University of Nigeria, Carver Building, 1 University Road, Nsukka 410001, Nigeria}
\email{xxx}

\author{Prasanta Gorai}
\affiliation{Rosseland Centre for Solar Physics, University of Oslo, PO Box 1029 Blindern, 0315 Oslo, Norway}
\affiliation{Institute of Theoretical Astrophysics, University of Oslo, PO Box 1029 Blindern, 0315 Oslo, Norway}
\email{xxx}

\author{kee-tae Kim}
\affiliation{Korea Astronomy and Space Science Institute, 776 Daedeokdae-ro, Yuseong-gu, Daejeon 34055, Republic of Korea}
\affiliation{University of Science and Technology, Korea (UST), 217 Gajeong-ro, Yuseong-gu, Daejeon 34113, Republic of Korea}
\email{xxx}

\author{Chang Won Lee}
\affiliation{Korea Astronomy and Space Science Institute, 776 Daedeokdae-ro, Yuseong-gu, Daejeon 34055, Republic of Korea}
\affiliation{University of Science and Technology, Korea (UST), 217 Gajeong-ro, Yuseong-gu, Daejeon 34113, Republic of Korea}
\email{xxx}

\author{Tapas Baug}
\affiliation{S. N. Bose National Centre for Basic Sciences, Block-JD, Sector-III, Salt Lake City, Kolkata 700106, India}
\email{xxx}

\author{Swagat Ranjan Das}
\affiliation{Departamento de Astronom\'ia, Universidad de Chile, Casilla 36-D, Santiago, Chile}
\email{xxx}

\author[0000-0002-8614-0025]{Shivani Gupta}
\affiliation{Indian Institute of Astrophysics, Koramangala II Block, Bangalore 560034, India}
\affiliation{Pondicherry University, R.V. Nagar, Kalapet, 605014, Puducherry, India}
\email{xxx}

\author[0000-0001-7866-2686]{Jihye Hwang}
\affil{Institute for Advanced Study, Kyushu University, Japan}
\affil{Department of Earth and Planetary Sciences, Faculty of Science, Kyushu University, Nishi-ku, Fukuoka 819-0395, Japan}
\email{xxx}

\author{Leonardo Bronfman}
\affiliation{Departamento de Astronom\'ia, Universidad de Chile, Casilla 36-D, Santiago, Chile}
\email{xxx}

\author{Archana Soam}
\affiliation{Indian Institute of Astrophysics, Koramangala II Block, Bangalore 560034, India}
\email{xxx}

\author{L. K. Dewangan}
\affiliation{Astronomy $\&$ Astrophysics Division, Physical Research Laboratory, Navrangpura, Ahmedabad 380009, India}
\email{xxx}




\begin{abstract}
We present a systematic comparative analysis of 324 starless cores in early-phase infrared-dark clouds (IRDCs; ASHES survey) and evolved-phase infrared-bright clouds (IRBCs; QUARKS survey) using 1.3\,mm continuum and line data by the Atacama Large Millimeter/submillimeter Array (ALMA). Despite having comparable sizes (\(\sim 2500\) au), starless cores in IRBCs exhibit systematically higher median mass (\(1.5\,M_{\odot}\) vs. \(0.6\,M_{\odot}\)), number density, and surface density—enhancements of approximately a factor of two relative to starless cores in IRDCs. Starless cores in IRBCs also display relatively stronger non-thermal motions (\(\sigma \sim 0.5\)\,km\,s\(^{-1}\)  vs. \(0.3\)\,km\,s\(^{-1}\)), higher total virial parameters (median \(\alpha_{\mathrm{vir,tot}} \sim 2.3\) vs. \(1.0\)), and steeper density profiles, indicating more centrally concentrated structures in feedback-driven, turbulence-enhanced environments. These findings support a dual evolutionary origin: (i) new core formation in evolved IRBCs under altered initial conditions, and (ii) subsequent dynamical mass growth via accretion from extended reservoirs. The prevalence of low-mass starless cores—even in late-stage IRBC environments—challenges models requiring massive prestellar cores and instead favors competitive-like dynamical mass accretion scenarios for high-mass star formation.

\end{abstract}



\section{Introduction}\label{sec:intro}
The formation mechanism of high-mass stars (\(M_* \gtrsim 8\,M_{\odot}\)) remains one of the most challenging problems in modern astrophysics, despite decades of intensive research. Their formation in distant, clustered environments, rapid evolution, and intense feedback processes (e.g., radiation pressure, outflows, H\,\textsc{ii} regions) severely complicate observational investigations \citep{2003ARA&A..41...57L,2007ARA&A..45..481Z,2009ApJ...696..268Z,2018ARA&A..56...41M,2025ARA&A..63....1B}. It is widely accepted that high-mass stars originate within dense (\(n \sim 10^{6}\,\mathrm{cm}^{-3}\)), compact (\(\lesssim 0.1\) pc) structures known as dense cores, which themselves reside within larger (\(\sim 0.5\)–1 pc) molecular clumps \citep{2007ARA&A..45..339B,2015ApJ...804..141Z,2018ARA&A..56...41M}. Among these, starless cores—dense cores lacking internal protostars—represent the earliest observable stage of star formation. A subset of these, if gravitationally bound, are termed prestellar cores \citep{1994MNRAS.268..276W,2007prpl.conf...33W,2014prpl.conf...27A}, and are direct progenitors of future stars.

Two mainstream theoretical frameworks attempt to explain high-mass star formation: the turbulent core accretion model \citep{2003ApJ...585..850M} and the competitive accretion model \citep{2001MNRAS.323..785B}. The former requires the prior existence of massive prestellar cores (above a few \(10 \times \,M_{\odot}\)) that collapse monolithically to form a single high-mass star. In contrast, the latter posits that clumps first fragment into numerous low-mass, thermal-Jeans-scale cores, which then grow competitively by accreting gas from a common reservoir, likely via filamentary networks \citep{2009MNRAS.400.1775S,2019MNRAS.490.3061V,2020ApJ...900...82P}. Critically, distinguishing between these scenarios hinges on the observational identification of genuine high-mass prestellar cores \citep{2009ApJ...696..268Z,2013ApJ...779...96T,2018ARA&A..56...41M}.

However, despite extensive searches over the past two decades, confirmed high-mass prestellar cores remain exceedingly rare \citep{2014MNRAS.439.3275W,2017ApJ...849...25L,2023A&A...675A..53B,2024ApJ...961L..35M,2025A&A...696A..11V,2025ApJ...995..193Y}. The vast majority of observed starless cores—even in high-mass star-forming regions—are low-mass (\(M \lesssim 10\,M_{\odot}\)). This observational reality necessitates a paradigm shift: rather than focusing solely on elusive massive cores, we must systematically study the properties of low-mass starless cores within high-mass star-forming regions to understand how high-mass stars assemble their mass over time. While such cores have been extensively characterized in low-mass star-forming regions \citep{2008ApJ...684.1240E}, their counterparts in high-mass regions lack a comprehensive statistical analysis particularly across evolutionary stages, hindering our understanding of high-mass star formation.

Here, we leveraged high-resolution 1.3 mm continuum and spectral line data from two complementary surveys of 
the Atacama Large Millimeter/submillimeter Array (ALMA) to conduct the first systematic comparison of starless cores across the evolutionary sequence of high-mass star formation. 
The two surveys include the ALMA Survey of 70 \(\mu\)m-dark High-mass clumps in Early Stages (ASHES; targeting quiescent infrared-dark clouds (IRDCs) environments; \citealt{2019ApJ...886..102S,2022ApJ...939..102L,2023ApJ...949..109L}) and the QUARKS survey (Querying Underlying mechanisms of massive star formation with ALMA-Resolved gas Kinematics and Structures; targeting evolved infrared-bright clouds (IRBCs) environments; \citealt{2024RAA....24b5009L,2024RAA....24f5011X,2025ApJS..280...33Y}). Here, IRDCs and IRBCs represent approximately early and late stages of the evolutionary sequence for star formation, respectively.  IRDCs exhibit little to no ongoing activity, which preserves their dark appearance at infrared wavelengths (e.g., 8\,\um\,, 24\,\um\,, and 70\,\um\,; \citealt{2006ApJ...641..389R,2017ApJ...841...97S,2019ApJ...886..102S}). In contrast, IRBCs tend to be associated with active high-mass star formation, hosting strong feedback signatures such as molecular outflows and H\,\textsc{ii} regions that produce prominent infrared features (e.g.,  extended green objects, EGOs; \citealt{2008AJ....136.2391C} and photodissociation regions, PDRs; \citealt{2013PASA...30...57J,2015ApJ...815..130G,2016PASA...33...30R}).

\begin{table*}[!thb]
\centering
\caption{Parameters of targeted clumps and observing setups in both QUARKS and ASHES surveys  \label{tab:almaobs}}
\setcellgapes{1.2pt}
\renewcommand{\arraystretch}{1.8} 
\begin{tabular}{cccccccccc}
\hline
\hline
Surveys & $N_{\rm Clumps}$& Distances & Masses& Luminosities &FoV & MRS & Ang. Res.$^b$&  Mass Sen.$^c$  &$N_{\rm starless}$\\
& & (kpc) & (\msun) & (\lsun) & (\arcsec)&(\arcsec)&(\arcsec)&(\msun) & \\
(1) & (2) & (3) & (4) & (5) & (6) & (7) & (8) &(9)&(10)\\
\hline
QUARKS$^a$ & 139  &  $1.2-11.6 $  &   $10^2-10^5$   &   $10^3-10^7$     & $\sim$40&    $\sim$29& $\sim$1.2   &     $\sim$\,0.14 &127\\
ASHES Pilot & 12  &  $2.9-5.4$   &   $10^2-10^3$   &   $10-10^3$     &    $\sim$60 &    $\sim$29   & $\sim$1.2   &  $\sim$\,0.10 &197\\
\hline
\end{tabular}{
\begin{flushleft}
$^a$ QUARKS ACA 7-m+TM2 12-m combined data.\\
$^b$ Angular Resolution.\\
$^c$ Mass Sensitivity.\\
\end{flushleft}
}
\end{table*}

This paper is aimed to explore core formation mechanisms, evaluate the influence of environmental evolution, and ultimately shed light on the physical pathways governing stellar mass growth in high-mass star-forming regions. To do so, we  analyze the masses, densities, kinematics, and dynamical states of starless cores in both IRDC and IRBC environments,
The paper is organized as follows: Section\,\ref{sec:2} describes  the ALMA data from both surveys and the sample of candidate starless cores. Section\,\ref{sec:res} presents the results and analysis. Section\,\ref{sec:4} discusses the formation and evolution of starless cores. Section\,\ref{sec:5} summarizes the major results.

\section{ALMA Data and Sample of candidate starless cores}\label{sec:2}

\subsection{ALMA Data}
\label{sec:alma_data}

This study utilizes high-resolution ALMA 1.3\,mm observations from two complementary surveys, including the ASHES pilot survey focusing on IRDC environments and the QUARKS survey targeting IRBC environments. The combination of these datasets enables a comparative analysis of starless cores across distinct star-forming environments, facilitated by their well-matched observational configurations (see below).

The ASHES pilot survey (PI: Sanhueza; \citealt{2019ApJ...886..102S}) observed 12 massive, 70 \(\mu\)m-dark clumps in IRDCs with typical sizes of \(\sim\,1\)\,pc, masses of \(\sim10^{2}\)–\(10^{3}\,M_{\odot}\), bolometric luminosities of $\sim$10–\(10^{3}\,L_{\odot}\), luminosity-to-mass ratio ($L_{\rm bol}/M$) of $\sim0.05-1.7$\,\lsun/\msun\,, and distances of 2.9–5.4 kpc \citep{2019ApJ...886..102S,2022ApJ...939..102L,2023ApJ...949..109L}. Mosaic observations at 1.3\,mm combined the 12-m array, ACA 7-m array, and total power (TP). The combined 12-m and ACA 7-m continuum data have an MRS of $\sim$29\,\arcsec\,, a mosaic FoV of \(\sim 1^{\prime}\) (1.5 times larger than that in QUARKS), a synthesized beam of \(\sim 1.2^{\prime\prime}\), and a continuum sensitivity of \(\sim 0.1\) mJy beam\(^{-1}\). 
General information about the survey is summarized in Table\,\ref{tab:almaobs}.
Line data from the 12-m and ACA arrays were combined with TP, covering at least 10 transitions including N\(_2\)D\(^+\) \(J=3\)–2, DCN \(J=3\)–2, and CO \(J=2\)–1, with rms values of 9.5\,mJy\,beam\(^{-1}\) and 3.5\,mJy\,beam\(^{-1}\) at velocity resolutions of 0.17\,km\,s\(^{-1}\) and 1.3\,km\,s\(^{-1}\), respectively.
Detailed information on the spectral windows of the survey is listed in Table\,\ref{tab:almaspw}.

The QUARKS survey (PIs: Lei Zhu, Guido Garay, Tie Liu; Project ID: 2021.1.00095.S; \citealt{2024RAA....24b5009L,2024RAA....24f5011X,2025ApJS..280...33Y}) is a follow-up program to the ATOMS survey \citep{2020MNRAS.496.2790L,2021MNRAS.505.2801L} that provides higher angular resolution (\(\sim 0.3^{\prime\prime}\) versus \(2^{\prime\prime}\)) at 1.3\,mm in ALMA Band\,6. The survey targeted 139 massive protocluster clumps in IRBCs, with sizes of \(\sim\,1.5\) pc, masses of \(\sim10^{2}\)–\(10^{5}\,M_{\odot}\), luminosities of \(\sim10^{3}\)–\(10^{7}\,L_{\odot}\), $L_{\rm bol}/M$ of $\sim4-400$\,\lsun/\msun\,, and distances of 1–12 kpc. To map extended filamentary structures, 17 clumps were observed with two pointings, resulting in 156 fields. Observations employed three configurations: ACA 7-m array (\(\sim 5^{\prime\prime}\) resolution), 12-m compact array C-2/C-3 (TM2; \(\sim 1^{\prime\prime}\)), and extended C-5/C-6 (TM1; \(\sim 0.3^{\prime\prime}\)) (see \citealt{2024RAA....24b5009L}, Table 1). The spectral setup covered a 8\,GHz frequency bandwidth at a velocity resolution of \(\sim 1.3\) km s\(^{-1}\), including cold gas tracers (e.g., N\(_2\)D\(^+\) 3–2), outflow tracers (e.g., CO 2–1, SiO 5–4), hot molecular core tracers (CH\(_3\)CN 12–11), and the H30\(\alpha\) ionized gas tracer. In this study, we mainly used the combined TM2 and ACA 7-m data \citep{2025ApJS..280...33Y}, which achieved maximum recoverable scales (MRSs) of 20\,\arcsec\,–29\,\arcsec\,, a field of view (FoV) of \(\sim 40^{\prime\prime}\), a synthesized beam of \(\sim 1.2^{\prime\prime}\), and a continuum sensitivity of \(\sim 0.6\) mJy beam\(^{-1}\). 
The line data have an average rms of \(\sim 10\) mJy beam\(^{-1}\) per channel at a synthesized beam of ($\sim 1.2^{\prime\prime}$, see Table\,\ref{tab:almaspw}).

The ASHES and QUARKS datasets are well-suited for comparative analysis due to their closely matched observational parameters, which are summarized in Table\,\ref{tab:almaobs}. Both surveys achieved similar MRSs (\(\sim 29^{\prime\prime}\)) and synthesized beams (\(\sim 1.2^{\prime\prime}\)), ensuring consistent spatial filtering and resolution. Their continuum sensitivities (QUARKS: \(\sim 0.4\) mJy beam\(^{-1}\); ASHES: \(\sim 0.1\) mJy beam\(^{-1}\)) translate to comparable mass sensitivities of \(\sim 0.1\,M_{\odot}\) at typical distances of 4 kpc, accounting for environmental temperature differences (see Section \ref{sec:sample}). Note that the QUARKS continuum sensitivity of \(\sim 0.4\) mJy beam\(^{-1}\) is adopted here, as it applies specifically to the fields containing candidate starless cores (see below). 
This consistency in sensitivity, resolution, and spatial scale recovery significantly enhances the reliability of comparing starless core properties across the two surveys. In addition, the differing $L_{\rm bol}/M$ ratios between ASHES and QUARKS clumps that host dense cores, as characterized by median values of 0.3\,\lsun/\msun\, and 40\,\lsun/\msun\,, respectively, indicate that the starless cores in the two samples reside in star-forming environments at distinct evolutionary stages. This contrast enables a robust comparative analysis across the evolutionary sequence of high-mass star formation.

\begin{table*}[!thb]
\centering
\caption{Spectral Windows of two surveys  \label{tab:almaspw}}
\setcellgapes{1.2pt}
\renewcommand{\arraystretch}{1.8}
\setlength{\tabcolsep}{3pt}  
\begin{tabular}{ccccccc}
\hline
\hline
Surveys & Center Frequency & Bandwidth & Vel. Res.$^a$ &Ang. Res.$^b$& Line Sens.$^c$ & Note$^d$ \\
& (GHz) & (GHz) & ($\rm km\,s^{-1}$) & (\arcsec)& ($\rm mJy\,beam^{-1}$) \\
(1) & (2) & (3) & (4) & (5) & (6) &(7)\\
\hline
QUARKS&\makecell[c]{$\sim$$[217.918$, $220.319,$\\\\$\,\,\,\,231.370$, $233.520]$} & 1.875 & $\sim1.3$ &$\sim1.2$& $\sim10$ & 
\makecell[c]{$\rm CO\,(2-1)$, $\rm SiO\,(5-4)$\\ H$_2$CO (3$_{0,3}$–2$_{0,2}$),\\ H$_2$CO (3$_{2,2}$–2$_{2,1}$), \\H$_2$CO (3$_{2,1}$–2$_{2,0}$),\\ $\rm CH_3OH\,(4-3)$, \\$\rm CH_3CN\,(12-11)$, H30$\alpha$} \\
\hline
ASHES&\makecell[c]{$\sim$[216.087, 216.348,\\ \,\,\,217.079, 217.213] }&0.06&$\sim$0.07&$\sim1.2$&$\sim$9.5 &\makecell[c]{$\rm SiO\,(5-4)$}\\
&\makecell[c]{$\sim$[218.825, 230.831]}&1.875&$\sim$1.3&$\sim1.2$&$\sim$3.5&\makecell[c]{H$_2$CO (3$_{0,3}$–2$_{0,2}$),\\ H$_2$CO (3$_{2,2}$–2$_{2,1}$), \\H$_2$CO (3$_{2,1}$–2$_{2,0}$),\\ $\rm CH_3OH\,(4-3)$}\\
&\makecell[c]{$\sim$[231.193, 231.295]}&0.06&$\sim$0.17&$\sim1.2$&$\sim$9.5&\makecell[c]{$\rm CO\,(2-1)$}\\
\hline
\end{tabular}
{
\begin{flushleft}
$^a$ Velocity Resolution.\\
$^b$ Angular Resolution.\\
$^c$ Line Sensitivity. \\
$^d$ The primary molecules used to trace protostars in the two surveys, which are described in Section\,\ref{sec:sample} (see also \citealt{2019ApJ...886..102S} and \citealt{2025ApJS..280...33Y} for the ASHES and QUARKS survey, respectively). Among them, CO and SiO are employed to trace molecular outflows.\\
\end{flushleft}
}
\end{table*}

\subsection{Sample of Starless Cores}
\label{sec:sample}

Our sample of starless cores was drawn from the ASHES and QUARKS surveys, selected based on the absence of star formation signatures in molecular line emission.

The ASHES survey identified 294 dense cores, with 197 classified as starless candidates based on the absence of warm gas or outflow signatures \citep{2019ApJ...886..102S,2023ApJ...949..109L}. Given the cold IRDC environment, these cores are assumed to have typical temperatures of \(\sim 10\)\,K \citep{2013ApJ...773..123S,2023A&A...675A..53B,2024ApJ...961L..35M}. The continuum sensitivity of \(\sim 0.1\) mJy beam\(^{-1}\) yields a mass sensitivity of \(\sim 0.1\,M_{\odot}\) at an average distance of 4\,kpc, comparable to the QUARKS sample. It is worth noting that 
although the  ASHES complete survey observed additional 27 massive IRDC clumps, and yield 297 more candidate starless cores within \citep{2023ApJ...950..148M,2024ApJ...966..171M,2026ApJ...997..155M}, the lack of kinematic analysis (e.g., virial parameters, and gas velocity dispersion) on them in the literature limits the present study to the subset (i.e., 197 candidate starless cores) well-characterized by \cite{2019ApJ...886..102S} and \cite{2022ApJ...939..102L,2023ApJ...949..109L}.

From the QUARKS TM2 and ACA 7-m combined data, \cite{2025ApJS..280...33Y} identified 1,562 dense cores, classifying 127 as starless core candidates based on the lack of outflow, hot core, or ionized gas tracers. These cores reside in 60 of the 139 targeted IRBC clumps. The continuum sensitivity for this subset of targeted clumps is improved to be \(\sim 0.4\) mJy beam\(^{-1}\), corresponding to a mass sensitivity of \(\sim 0.14\,M_{\odot}\) at an average distance of 4\,kpc and a typical starless core temperature of 20\,K within IRBC environments (e.g., \citealt{2018A&A...618L...5N,2025A&A...696A..11V,2025ApJS..280...33Y}). This enhanced sensitivity enables robust detection of low-mass starless cores in IRBCs from the QUARKS survey, comparable to the ASHES survey (see above).

We stress that the criteria used to identify starless cores were consistent between the ASHES and QUARKS surveys due to their highly similar observing setups. Specifically, any source exhibiting signatures of outflows (as traced by broad CO and SiO emission), warm/hot molecular gas (e.g., CH$_3$OH (4–3), H$_2$CO (3$_{2,2}$–2$_{2,1}$/3$_{2,1}$–2$_{2,0}$), CH$_3$CN), or ionized gas (e.g., H30$\alpha$) was classified as protostellar or evolved and excluded from the starless core sample \citep{2019ApJ...886..102S,2025ApJS..280...33Y}. We finally only retained as candidate starless cores those detected in molecular lines with upper energy levels $E_u < 22\,\mathrm{K}$ (e.g., N$_2$D$^+$ (3–2), DCN (3–2), CO/$^{13}$CO/C$^{18}$O (2–1), and H$_2$CO (3$_{0,3}$–2$_{0,2}$)). 

The combined sample of 324 candidate starless cores (197 from ASHES, 127 from QUARKS) provides an  opportunity to study core properties across IRDC and IRBC environments, with consistent mass sensitivities (\(\sim 0.1\,M_{\odot}\)) minimizing bias in comparative analysis. It is worth noting that the candidate starless cores (for simplicity, hereafter called starless cores)  from the two surveys were identified using different source extraction algorithms. The potential influence of these methodological differences on comparative analyses will be discussed elsewhere as needed.

 \begin{figure*}[ht!]
    \centering
    \includegraphics[angle=0, width=1\textwidth]{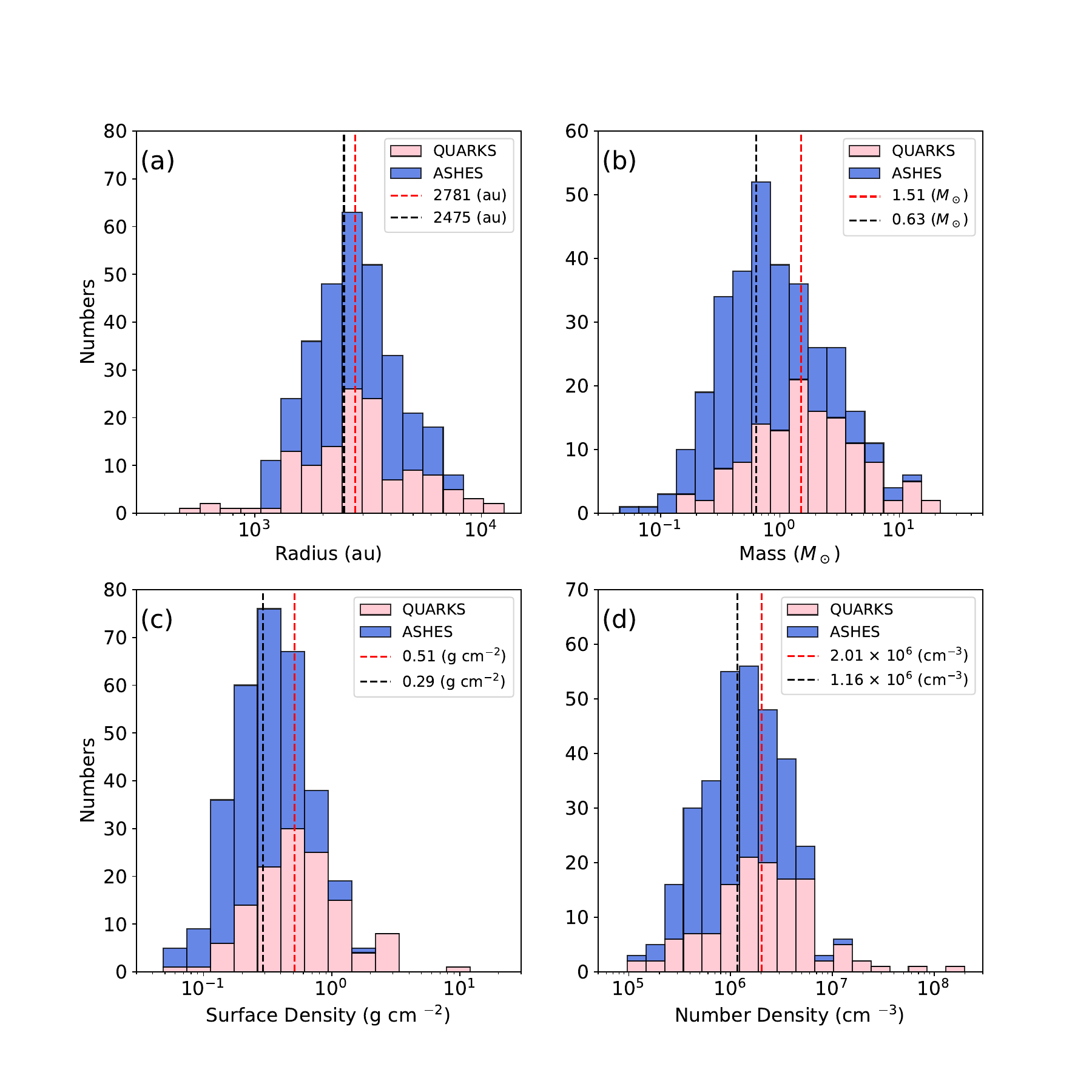}  
    \caption{Distribution of the physical parameters of starless cores from the ASHES and QUARKS surveys, including  the radius (panel\,a), mass (panel\,b), surface density (panel\,c), and number density (panel\,d). The black and red dashed lines in each panel represent the median values for the ASHES and QUARKS samples, respectively.} 
    \label{fig:pro}    
\end{figure*}

\section{Results}\label{sec:res}

 \begin{figure*}[ht!]
    \centering
    \includegraphics[angle=0, width=0.95\textwidth]{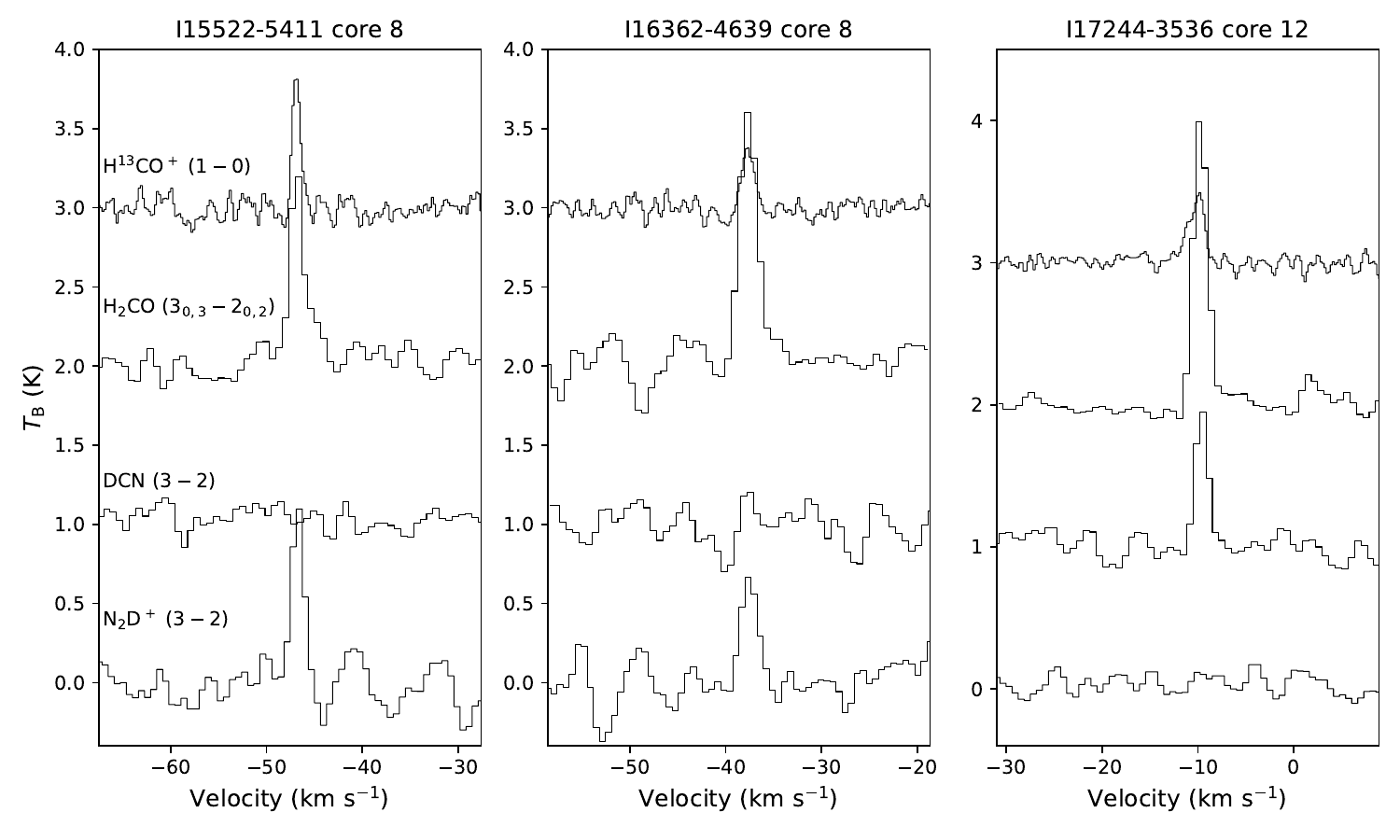} 
    \caption{Example of core-averaged spectra of molecular lines from the QUARKS sample. The transitions of $\rm H^{13}CO^+$, $\rm H_2CO$, DCN, and $\rm N_2D^+$ are indicated in the first panel} 
    \label{fig:spectrum}    
\end{figure*}

\subsection{Physical parameters of starless cores}\label{sec:3.2}
We retrieved fundamental parameters of starless cores—including radius, mass, mass surface density, and number density—from the ASHES \citep{2019ApJ...886..102S,2023ApJ...949..109L} and QUARKS \citep{2025ApJS..280...33Y} catalogs. For the QUARKS sample, we primarily used the full catalog of 127 candidate starless cores in our analysis and subsequent discussion. However, we note that this catalog may contain a small fraction of protostellar contaminants; the potential impact of such sources on our results is assessed in detail in Sect.\,\ref{sec:limit}. In the following, we examine several key uncertainties that may affect the parameter estimates in both surveys.

The two surveys employed distinct source extraction algorithms in identifying dense cores: \texttt{astrodendro} for ASHES and \texttt{getsf} for QUARKS.
Previous studies indicate that \texttt {getsf} tends to favor spherical structures but could miss irregular ones \citep{2024ApJS..270....9X,2024ApJ...967...56C}.
This leads to a direct consequence of the detection number of density structures. For instance, when \texttt{getsf} was applied to the ASHES data, it identified only 66 cores compared to 297 with \texttt{astrodendro} \citep{2024ApJS..270....9X}. 
However, the influence of this  methodological difference on systematic analysis of starless core properties investigated here could be ignored (see below, and Appendix\,\ref{app:c}).

Temperature assumptions also differed: ASHES starless core temperatures were measured via NH\(_3\) (1,1)/(2,2) line ratios, yielding \(T_{\text{NH}_3} \sim 7\)–22.8\,K (median $\sim$14\,K; \citealt{2023ApJ...949..109L}), consistent with the cold IRDC environment. QUARKS adopted a fixed \(T_{\text{dust}} = 20\)\,K, typical for starless cores in clustered IRBC regions \citep{2013MNRAS.432.3288S}. This temperature difference could introduce systematic biases in mass estimates and thus in following systematic analysis, as \(M \propto 1/T_{\text{dust}}\) under optically thin assumptions, which will be discussed below.

 Despite these methodological differences, the starless cores exhibit similar radius distributions (Figure \ref{fig:pro}a). The ASHES starless cores span \(1200\)–\(7000\)\,au (median \(2500\)\,au), while QUARKS ones range \(500\)–\(10000\)\,au (median \(2800\)\,au). A two-sided Kolmogorov–Smirnov (K-S) test\footnote{The Kolmogorov–Smirnov (K-S) test assesses whether two samples originate from the same distribution by calculating a $p$-value; a value below 0.05 typically rejects this null hypothesis. A two-sided test detects any significant difference between distributions, while a one-sided test evaluates a specific directional difference (e.g., whether one sample is stochastically larger than the other).}  yields \(p \sim 0.11\), indicating the radii are statistically consistent with being drawn from the same parent distribution. This suggests we are detecting density structures on comparable physical scales, and that methodological differences in source extraction do not systematically bias radius measurements. 
 
Figure \ref{fig:pro}b–d compares the mass, mass surface density, and number density distributions between the two samples. Starless cores within IRBCs (QUARKS sample) exhibit a median mass of 1.51 \(M_{\odot}\), approximately 2.4 times greater than those in IRDCs (ASHES sample; 0.63 \(M_{\odot}\)). This result aligns with \cite{2024ApJS..270....9X}. A two-sided K-S test yields \(p \ll 0.05\), indicating the mass distributions are drawn from statistically distinct parent populations, and a one-sided test confirms QUARKS cores are systematically more massive (\(p \ll 0.05\)). Similarly, QUARKS cores show significantly greater mass surface density and number density than ASHES cores (Figure\,\ref{fig:pro}c-d; both K-S tests yield \(p \ll 0.05\)).

Furthermore, we evaluated the robustness of these differences against three potential biases (see Appendices \ref{app:a} and \ref{app:b} for details). First, limiting both samples to a common distance range (2.9–5.4\,kpc) did not alter the mass difference. Second, comparing 127 randomly selected QUARKS starless cores with 127 certain ASHES starless cores preserved the mass difference. Third, Monte Carlo (MC) simulations with randomized \(T_{\text{dust}}\) (7–22.8\,K for ASHES; 10–30\,K for QUARKS) confirmed temperature uncertainties cannot account for the observed mass difference. 
These evaluations collectively demonstrate that the observed differences in core properties are robust against methodological and observational biases. The systematic enhancement in mass, number density, and surface density of QUARKS starless cores relative to ASHES cores could reflect actual environmental differences between IRBCs and IRDCs, rather than analysis artifacts.

 \begin{figure*}[ht!]
    \centering
    \includegraphics[angle=0, width=0.4\textwidth]{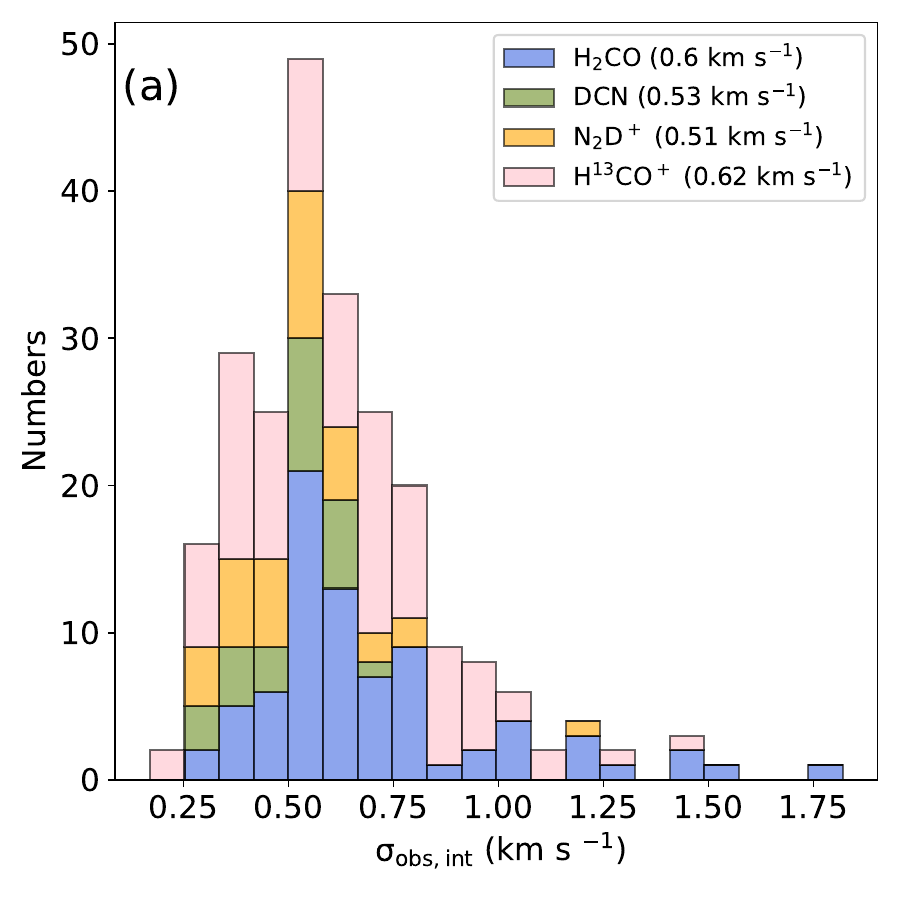} 
    \includegraphics[angle=0, width=0.4\textwidth]{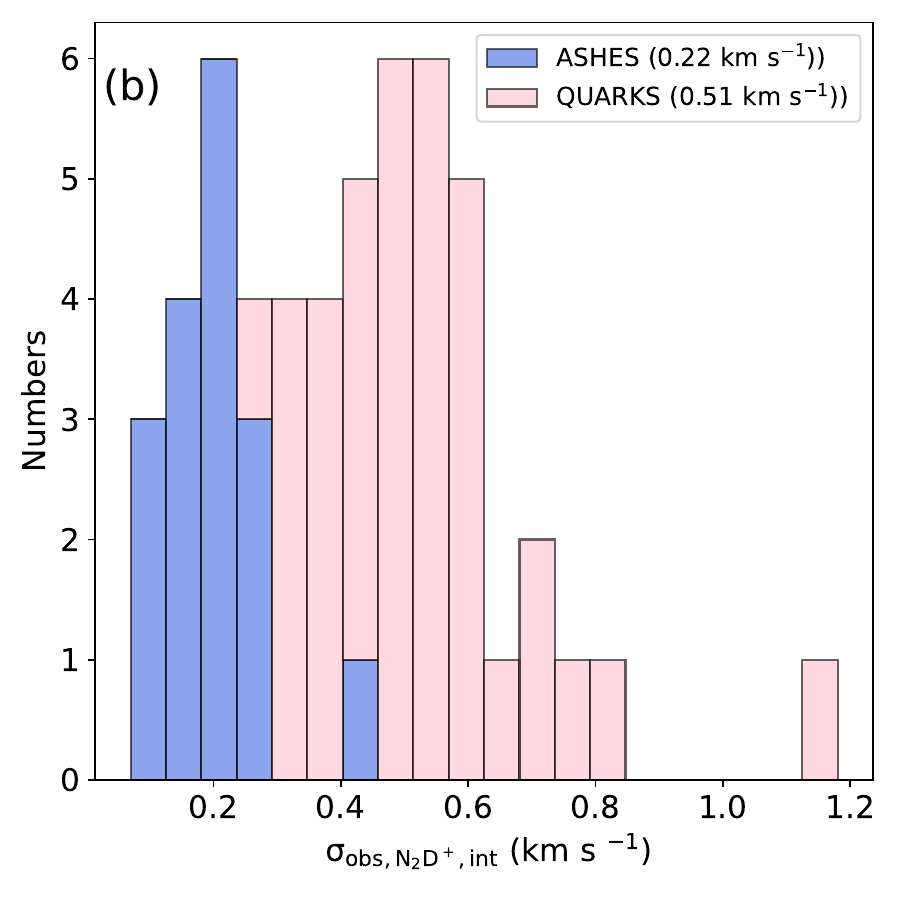}
    \caption{Panel\,(a): distribution of the intrinsic observed velocity dispersions (corrected by the velocity resolution, see Sect.\,\ref{subs:3.3}) in the QUARKS starless cores. The color-coding corresponds to different molecular tracers. Panel\,(b): distribution of the intrinsic observed velocity dispersion from the $\rm N_2D^+$ molecule between the ASHES (17) and QUARKS (36) starless core samples.
    The median value with the standard deviation for each group is provided in parentheses.
    } 
    \label{fig:line:vd}    
\end{figure*}

\subsection{Analysis of Molecular Line data}\label{subs:3.3}
To analyze the kinetic properties (e.g., virial analysis) of starless cores, it is necessary to characterize their associated gas velocity dispersion. For ASHES starless cores, \cite{2023ApJ...949..109L} measured their velocity dispersion of the observed $\rm DCO^+\,(3-2)$ and $\rm N_2D^+\,(3-2)$ lines. For QUARKS starless cores, we do the same measurement but using several molecular line emission, including $\rm H^{13}CO^+\,(1-0)$ line  from the ATOMS survey \citep{2020MNRAS.496.2790L,2021MNRAS.505.2801L} and $\rm H_2CO\,(3_{0,3}-2_{0,2})$, DCN\,$(3-2)$, and $\rm N_2D^+\,(3-2)$ line data from the QUARKS survey.

From the core-averaged spectra extracted for the 127 QUARKS starless cores, we find that $\rm H^{13}CO^+$ has the highest detection rate of  $\sim74.8\%$ (95/127), followed by $\rm H_2CO$ at $\sim61.4\%$ (78/127), $\rm N_2D^+$ at $\sim28.3\%$ (36/127), and DCN at $\sim20.5\%$ (26/127). Both $\rm N_2D^+$ and DCN emission were detected toward 8 QUARKS starless cores ($6\%$). In total, 54 cores ($42.5\%$) were detected in either $\rm N_2D^+$, DCN, or both.
As examples, Figure\,\ref{fig:spectrum} presents the spectral lines of these molecules for three selected QUARKS starless cores. Among the 127 cores, 23 ($\sim18.1\%$) show no detection in any of the four lines investigated here, while 104 ($\sim81.9\%$) cores have at least one molecular line detection. 

To characterize the line profiles, we fitted each line spectrum to Gaussian functions with the best-fit parameters—brightness temperature ($T_{B}$), central velocity ($V_{\rm LSR}$), and observed velocity dispersion ($\sigma_{\rm obs}$)—provided in Table\,\ref{table:line:pro} for each starless core and molecular line. It is noteworthy that while a multi-Gaussian fitting was applied to spectra with multiple velocity components, Table\,\ref{table:line:pro} only reports the parameters of the major component with the higher $T_{b}$. Subsequently, the intrinsic observed velocity dispersion follows $\sigma_{\rm obs,int}=\sqrt{\sigma_{\rm obs}^2-\rm FWHM^2/8ln2}$, where $\sigma_{\rm obs}$ is the observed velocity dispersion as shown in Table\,\ref{table:line:pro}, and $\rm FWHM$ is the velocity resolution ($\sim\rm 0.4\,km\,s^{-1}$ for $\rm H^{13}CO^+$ and $\sim\rm 1.3\,km\,s^{-1}$ for other molecules).

As shown in Figure\,\ref{fig:line:vd}, the $\rm H_2CO$ and $\rm H^{13}CO^+$ lines show slightly greater $\sigma_{\rm obs,int}$ (median $\rm \sim 0.6\,km\,s^{-1}$) than the $\rm N_2D^+$ and DCN lines (median $\rm\sim 0.5\, km\,s^{-1}$). This implies that the former pair of molecular lines primarily may probe the slightly outer envelope gas of the cores, whereas the latter pair traces relatively denser, inner regions. In comparison, the $\sigma_{\rm obs,int}$ of the QUARKS starless cores are overall ~1.6-2.0 times higher than those measured in the ASHES starless cores (median value $\sim\rm 0.3\,km\,s^{-1}$; \citealt{2023ApJ...949..109L}). However, we note that the ASHES measurements were based on $\rm DCO^+\,(3-2)$ and $\rm N_2D^+\,(3-2)$ emission. To ensure a robust comparison, we consider the common tracer only, $\rm N_2D^+$, between both surveys. A direct comparison of  $\sigma_{\rm obs, N_2D^+,int}$ for 17 ASHES and 36 QUARKS starless cores confirms the significant difference, with median value $\rm 0.51\,km\,s^{-1}$ vs. $\rm 0.22\,km\,s^{-1}$, respectively (Figure \ref{fig:line:vd}b).
The increase in $\sigma_{\rm obs,int}$ could stem from the higher turbulence expected in IRBC environments due to strong stellar feedback (e.g., outflows, HII regions, \citealt{2014prpl.conf..243K,2014prpl.conf..149T}) 
 than in IRDC environments due to quiet star formation.

\subsection{Stability of the dense cores}
\label{subsec:3.4}
The virial parameter ($\alpha_{\rm vir}$), which is commonly used to assess the gravitational stability of a dense core, is defined as follows \citep{1992ApJ...395..140B,2007ApJ...661..262D}:
\begin{equation}\label{eq:alpha}
    \alpha_{\rm vir}=\frac{5}{a\beta}\frac{\sigma^2_{\rm eff}R}{GM_c},
\end{equation}
where $a=(1-b/3)/(1-2b/5)$ is the correction factor for a power-law density profile ($\rho\propto R^{-b}$), and $\beta$ is the geometry factor. Following \citet{2013ApJ...768L...5L,2023ApJ...949..109L},  we adopted $\beta = 1.2$  and a typical density profile index of $b = 1.6$ (e.g., \citealt{2002ApJ...566..945B,2012ApJ...754....5B,2014ApJ...785...42P,2019A&A...622A..99L}).
$\sigma_{\rm eff}$ is the effective velocity dispersion,
$R$ is the radius of the core, G is the gravitational constant, and $M_c$ is the mass of the core.
Dense cores with $\alpha_{\rm{vir}} \leq 2$ are considered to be in a gravitationally bound state \citep{2013ApJ...779..185K,2022MNRAS.510.5009L}.

 \begin{figure}[ht!]
    \centering
    \includegraphics[angle=0, width=0.54\textwidth]{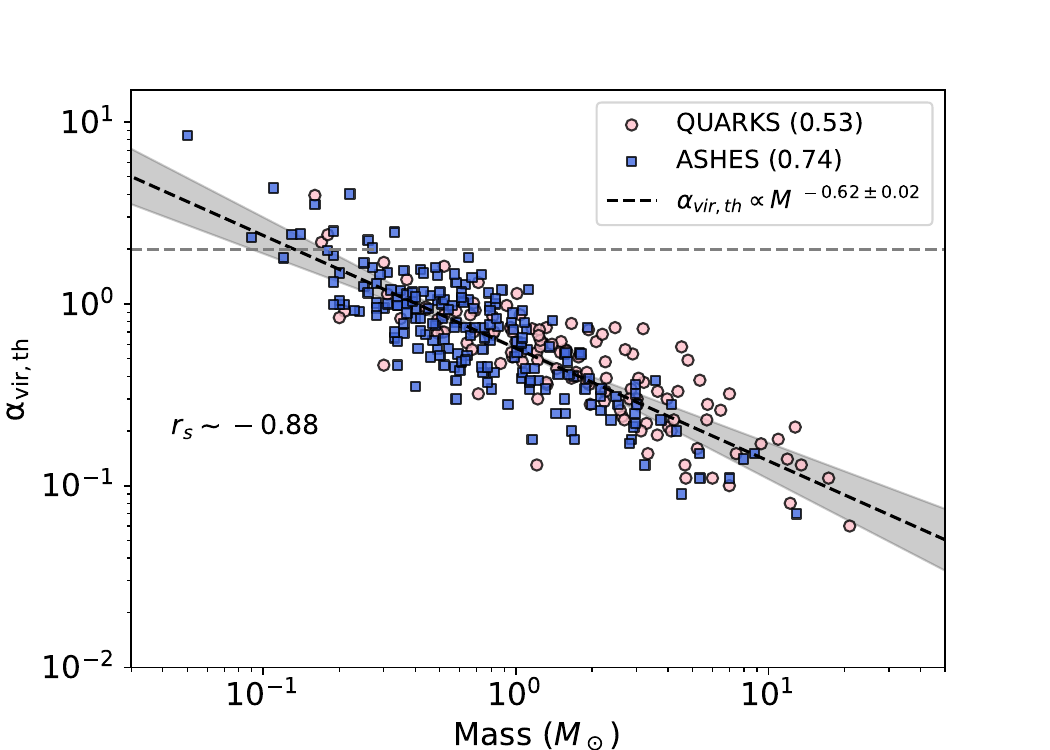} 
    \caption{Relation of $\alpha_{\rm vir,th}$ with  mass for starless cores from the ASHES (filled circles) and QURKS (filled squares) surveys.  The values in parentheses within the top-right legend correspond to the median values of $\alpha_{\rm vir,th}$. A linear regression fit to all sample points (black dashed line) yields a slope of $-0.62\pm0.02$. The Spearman’s rank test returns a coefficient of $r_s=-0.88$. The horizontal dashed line marks the critical value of the virial parameter equals 2.
    } 
    \label{fig:alpha:m}    
\end{figure}

\subsubsection{Thermal support}\label{subsub:3.4.1}
If only thermal motions are considered, the term $\sigma_{\rm eff}$ in Eq.\,\ref{eq:alpha} can be replaced by the sound speed of the gas, $c_{\rm s}=\sqrt{k_BT/\mu_p m_{H}}$, where $\mu_{p}=2.33$ is the mean molecular weight per free particle and $m_{\rm H}$ is the mass of the hydrogen atom \citep{2008A&A...487..993K}.
Under the assumption of a dust temperature of 20\,K, our analysis of the QUARKS 127 starless cores shows that the thermal virial parameter ($\alpha_{\rm vir,th}$) ranges from $\sim$0.1 to 4, with a mean value of $\sim$0.5—most 98\% (124/127) of which are below 2. Similarly, 94\% (186/197) of the ASHES starless cores have a $\alpha_{\rm vir,th}$ below 2 with a median value of $\sim$0.7. Take together, we  suggest that, when only the contribution of thermal motions is considered, most of both ASHES and QUARKS starless cores can be  gravitationally bound state.

Figure \ref{fig:alpha:m} presents the correlation between the thermal virial parameter (\(\alpha_{\rm vir,th}\)) and core mass for starless cores in both the ASHES and QUARKS surveys. A gradual decrease in \(\alpha_{\rm vir,th}\) with increasing mass is observed, characterized by a linear regression slope of \(\sim 0.62 \pm 0.02\) and a strong negative Spearman correlation coefficient (\(r_s \sim 0.88\)). This trend is consistent with the mathematical formulation in Equation \ref{eq:alpha}, where \(\alpha_{\rm vir}\) is inversely proportional to core mass. A similar trend was previously reported by \cite{2023ApJ...949..109L} for the full ASHES sample (containing both starless and prostellar cores), suggesting that more massive cores tend to be more gravitationally unstable.

Furthermore, we find from Figure \ref{fig:alpha:m}  that QUARKS starless cores in IRBC environments exhibit systematically lower virial parameters (median \(\alpha_{\rm vir,th} = 0.53\)) compared to ASHES cores in IRDC environments (median \(\alpha_{\rm vir,th} = 0.74\)). A one-sided K-S test confirms this difference is statistically significant (\(p \ll 0.05\)). This result implies enhanced gravitational instability in the QUARKS starless cores, rendering them more susceptible to collapse than their ASHES counterparts. The difference in gravitational stability between the two samples could arise from their distinct density profiles, as discussed in Section \ref{subsubsec:growth_scenario2}.

\subsubsection{ Turbulent support}\label{subsub:3.4.2}
We incorporated both thermal and turbulent motions by substituting the effective velocity dispersion in Eq.\,\ref{eq:alpha} with the total velocity dispersion, defined as \(\sigma_{\rm tot}^2 = \sigma_{\rm nt}^2 + c_s^2\). The non-thermal velocity dispersion was calculated as \(\sigma_{\rm nt} = (\sigma_{\rm obs, int}^2 - k_B T / m_{\rm obs})^{1/2}\), where \(m_{\rm obs}\) is the molecular mass of the tracer species (\(m_{\rm obs} = 30 m_{\rm H}\) for H\(_2\)CO, N\(_2\)D\(^+\), and H\(^{13}\)CO\(^+\); \(m_{\rm obs} = 28 m_{\rm H}\) for DCN).

 \begin{figure}[ht!]
    \centering
    \includegraphics[angle=0, width=0.47\textwidth]{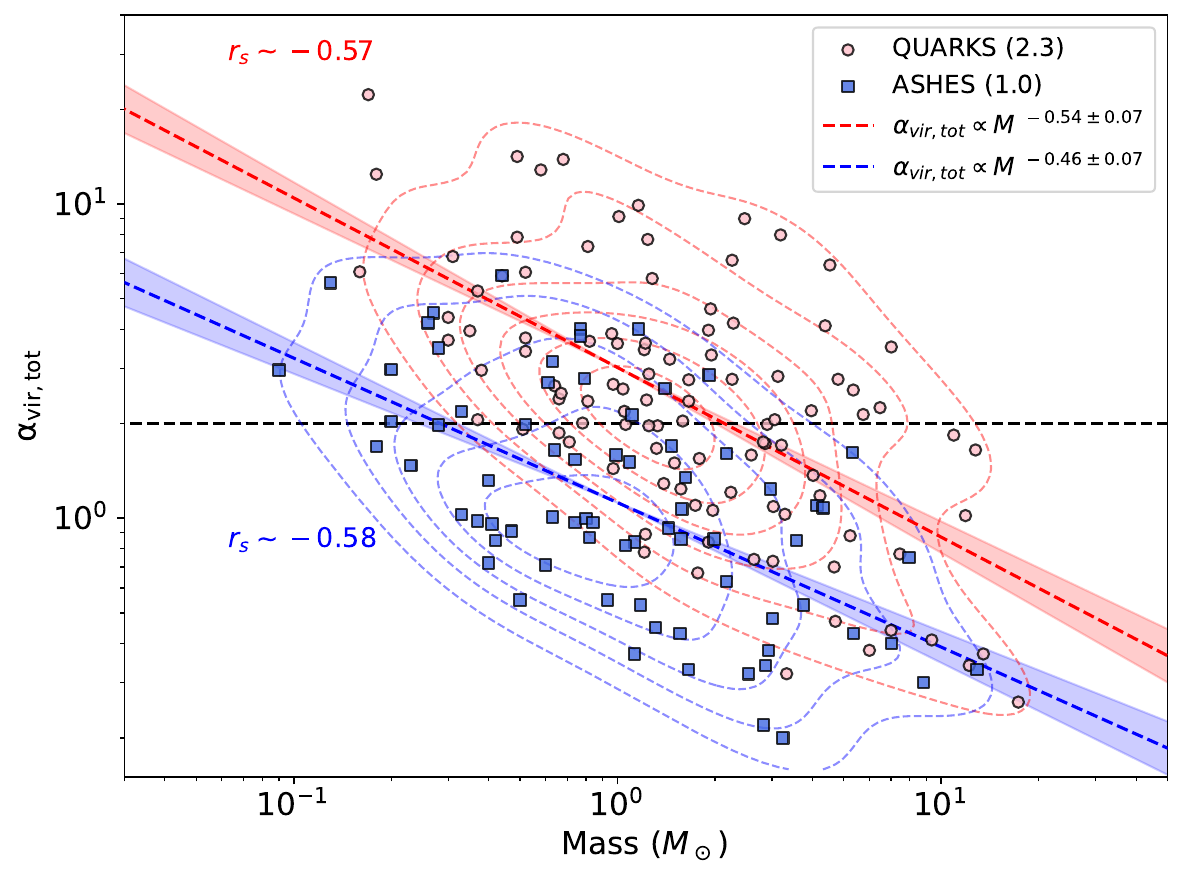} 
    \caption{Relation of $\alpha_{\rm vir,tot}$ with  mass for starless cores from the ASHES (filled circles) and QURKS (filled squares) surveys. The blue and red contours represent the Kernel Density Estimation distributions of the scatter points from the two survey samples, respectively. The red and blue dashed lines denote the linear regression fits to the two samples, yielding slopes of $\sim -0.5$ for both. The Spearman's rank test yielded correlation coefficients of $r_s=-0.57$ and $r_s=-0.58$, respectively. The horizontal dashed line corresponds to a critical value of  $\alpha_{\rm vir,tot}=2$. The values in parentheses within the top-right legend correspond to the median $\alpha_{\rm vir,tot}$ values.  } 
    \label{fig:alpha_tot:m}    
\end{figure}

As noted in Section \ref{subs:3.3}, 104 of the 127 QUARKS starless cores have detections of at least one of the aforementioned molecular lines, and thus only these cores are considered in the following analysis. For the QUARKS sample, we prioritized using N\(_2\)D\(^+\) and DCN emission to derive \(\sigma_{\rm nt}\), as these molecules preferentially trace the inner dense gas of cold cores. For cores lacking N\(_2\)D\(^+\) or DCN detections, we use H\(^{13}\)CO\(^+\) and H\(_2\)CO, which may trace relatively outer envelope gas. Note that the difference in probing depth between these tracers may not significant in the present study (Section \ref{subs:3.3}). For the ASHES sample, \cite{2023ApJ...949..109L} used DCO\(^+\) (3–2) and N\(_2\)D\(^+\) (3–2) lines to obtain \(\sigma_{\rm nt}\) for 73 starless cores with line detection above 3\(\sigma\).

After accounting for both thermal and turbulent support, we find that 44 (42.3\%) of the 104 QUARKS starless cores have total virial parameter (\(\alpha_{\rm vir,tot}\)) $<2$ and are thus potentially gravitationally bound, compared to 55 (75.3\%) of the 73 ASHES starless cores. 
Because different molecular tracers can probe distinct density regimes within prestellar cores \citep{2015A&A...579A..80G,2021ApJ...907L..15S,2023ApJ...945..156S}, we report here for readers as a reference the distribution of tracers used in estimating $\alpha_{\rm vir,tot}$, particularly for the QUARKS sample: 36 cores ($\sim$34.6\%) are traced by N$_2$D$^+$, 18 ($\sim$17.3\%) by DCN, 43 ($\sim$41.3\%) by H$^{13}$CO$^+$, and 7 ($\sim$6.7\%) by H$_2$CO.  Despite this tracer diversity, the QUARKS cores overall exhibit significantly higher median \(\alpha_{\rm vir,tot}\) values (\(\sim 2.3\)) than the ASHES cores (\(\sim 1.0\)), with a one-sided K-S test confirming this difference is statistically significant (\(p \ll 0.05\)). Even when restricting the comparison to cores detected in the same tracer—specifically N$_2$D$^+$ (36 in QUARKS and 17 in ASHES; see Figure~\ref{fig:line:vd}b)—the median $\alpha_{\rm vir,tot}$ remains much higher in QUARKS ($\sim$1.7) than in ASHES ($\sim$0.8). 
This can be attributed primarily to enhanced turbulence in infrared-bright cloud (IRBC) environments, which increases the derived \(\alpha_{\rm vir,tot}\) for starless cores therein.

We also examined in Figure \ref{fig:alpha_tot:m} the relationship between \(\alpha_{\rm vir,tot}\) and core mass (\(M_{\rm core}\)) for both samples. 
A tentative anti-correlation is observed, with more massive cores tending to have lower \(\alpha_{\rm vir,tot}\) values, consistent with the trend found when considering thermal support alone (see Figure \ref{fig:alpha:m}). This trend is quantified by power-law fits: \(\alpha_{\rm vir,tot} \propto M^{-0.54 \pm 0.07}\) for the QUARKS sample (red dashed line) and \(\alpha_{\rm vir,tot} \propto M^{-0.46 \pm 0.07}\) for the ASHES sample (blue dashed line). Spearman rank correlation coefficients (\(r_s \sim -0.57\) for QUARKS, \(r_s \sim -0.58\) for ASHES) indicate a moderate-strength inverse relationship. However, this global trend exhibits substantial intrinsic scatter where at any given mass, \(\alpha_{\rm vir,tot}\) spans nearly an order of magnitude (e.g., from \(\sim 0.3\) to \(\gtrsim 10\) for the QUARKS starless cores). This large dispersion suggests that mass alone is insufficient to determine a core’s dynamical state, and that other factors—such as the local environment, turbulence, and/or magnetic fields— could play roles in governing virial equilibrium. Apparently, the slope of the virial parameter–mass relation observed here could be in agreement with that found in numerical simulation results for turbulent and magnetized clouds (e.g., \citealt{2007ApJ...661..262D}). Nevertheless, the observational study of magnetic fields in early-stage cores remains challenging (e.g., \citealt{2025ApJ...980...87S}), and given the lack of observational constraints, we refrain from speculating in this direction.

It is worth noting that in our QUARKS sample, some very low-luminosity protostellar cores below the observing detection of the QUARKS survey may have been misclassified as starless sources (see below). Consequently, this misclassification could be artificially inflating the measured non-thermal velocity dispersions ($\sigma_{\rm nt}$) in the QUARKS sample, thereby skewing the virial analysis. Indeed, when scrutinizing the QUARKS sample using higher-resolution and sensitivity ALMA data, we find that about $21.3\%$ of the current QUARKS starless cores could be protostellar and thus misclassified (see more in Sect.\,4.3). The observed velocity dispersion of these misclassified starless cores is slightly higher than that of the remaining candidate starless cores in the QUARKS sample ($\sim\rm0.72\,km\,s^{-1}$ vs. $\sim\rm0.61\,km\,s^{-1}$, tracer: $\rm H^{13}CO^+$). Accordingly, this suggests that some misclassifications present in our QUARKS starless sample could slightly bias our virial parameter analysis, which should be cautioned.

\begin{figure}[ht!]
    \centering
    \includegraphics[angle=0, width=0.47\textwidth]{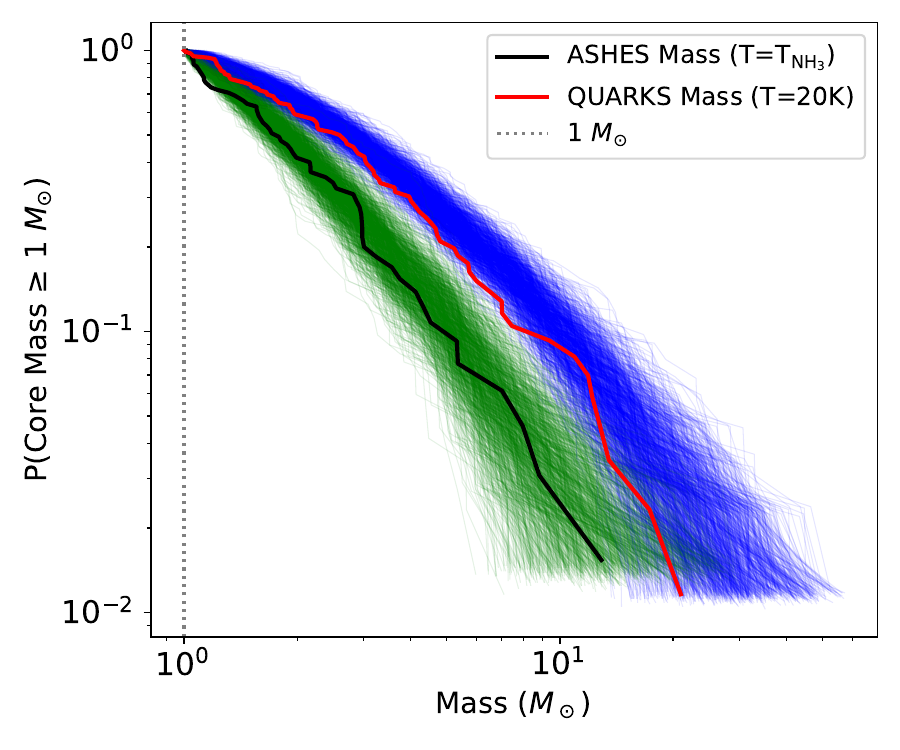} 
    \caption{Inverse cumulative distribution of the starless core mass in the ASHES and QUARKS samples.
    The green and blue bands represent the masses from 1000 simulations with two specific temperature ranges for the ASHES and QUARKS sample (see text), respectively.
    The red and black lines present the mass distributions derived from the individual $T_{\rm NH_3}$ measurement for the ASHES sample \citep{2023ApJ...949..109L} and from an assumed dust temperature of 20\,K for the QUARKS sample, respectively. The dashed line stands for a   threshold of \(M_{\mathrm{core}} > 1\,M_{\odot}\), corresponding to the peak of the mass distribution for all starless cores from both surveys (Figure \ref{fig:pro}b).
    } 
    \label{fig:cmf}    
\end{figure}
\subsection{Core mass function}\label{subsec:3.5}

We investigated the core mass function (CMF) to quantify differences in starless core masses between IRDC and IRBC environments. To mitigate biases inherent in differential CMF analysis—such as sensitivity to binning parameters and non-Gaussian noise in log-log space \citep{2009SIAMR..51..661C,2024A&A...690A..33L}, we adopted the inverse cumulative number density, \(P(M_{\mathrm{core}} \geq M)\), following \cite{2025A&A...696A.151C}. This cumulative representation is immune to binning choices and better preserves the high-mass tail \citep{2013A&A...551A.111O,2021ApJ...913...29F,2023MNRAS.525.4744P}. Prior to constructing the cumulative distributions, we assessed the impact of dust temperature uncertainties on mass estimates through 1000 Monte Carlo simulations, assigning random temperatures within observed ranges: 7.0–22.8 K for ASHES (based on NH\(_3\) measurements, $T_{\rm NH_3}$; \citealt{2023ApJ...949..109L}) and 10–30 K for QUARKS (accounting for warmer IRBC environments).

Figure \ref{fig:cmf} presents the inverse cumulative distribution for starless cores with \(M_{\mathrm{core}} > 1\,M_{\odot}\), a threshold corresponding to the peak of the mass distribution for all starless cores from both surveys (Figure \ref{fig:pro}b). The ASHES and QUARKS mass distributions from the 1000 temperature simulations are represented by green and blue bands, respectively. The black line shows ASHES core masses estimated from \(T_{\mathrm{NH_3}}\), while the red line represents QUARKS core masses assuming \(T = 20\,\mathrm{K}\). The cumulative distribution reveals a systematic shift: the QUARKS population consistently lies above the ASHES population, indicating a higher fraction of massive cores at any given mass threshold. For instance, at \(M = 2\,M_{\odot}\), \(P \sim 0.59\) for QUARKS versus \(\sim 0.41\) for ASHES; at \(M = 5\,M_{\odot}\), the probabilities drop to \(\sim 0.21\) and \(\sim 0.09\), respectively—demonstrating a clear enrichment of massive cores (\(\gtrsim 2\,M_{\odot}\)) in QUARKS. This systematic shift is robust against observational uncertainties, as the uncertainty bands remain separated even at their extremes, confirming the mass offset is not an artifact of temperature assumptions. 
In addition, an alternative artifact from the methodological 
difference on extraction of cores between the {\it astrodendro} algorithm for the ASHES survey and the {\it getsf} algorithm for the QUARKS survey can get ruled out since 
correcting for the influence of the core extraction algorithm will widen the mass shift between the ASHES and QUARKS samples (see Sects.\,4.3 and Appendix\,\ref{app:c}).



\section{Discussion}\label{sec:4}

\subsection{Formation and Evolution of Starless Cores}
\label{subsec:formation_evolution}

The comparative analysis reveals a systematic enhancement in the mass, number density, and surface density of starless cores in the QUARKS survey relative to the ASHES survey, with QUARKS cores exhibiting values approximately twice as high across all three parameters (see Figure\,\ref{fig:pro}b-d). This trend is corroborated by the cumulative core mass distribution, which shows the QUARKS population consistently above the ASHES population (see Figure\,\ref{fig:cmf}), indicating a higher prevalence of relatively massive cores. We interpret this systematic shift as a signature of evolutionary progression given that ASHES targets cold, quiescent IRDC clumps while QUARKS samples warmer, actively star-forming IRBC environments (extending to the ultra-compact H\,\textsc{ii} region phase).  More evolved clumps from QUARKS host more massive starless cores (see also Section~\ref{sec:3.2}), consistent with recent findings from the ALMAGAL survey \citep{2025A&A...696A.151C}. This raises a key question: through what physical mechanisms do starless cores acquire mass and density within natal clumps at different evolutionary stages? Two plausible scenarios are considered: (1) core growth over time via dynamical mass accretion, and (2) core formation under distinct initial conditions, where low-mass cores originate in cold, turbulence-quiescent IRDCs, while higher-mass cores form in warm, turbulence-active IRBCs.

\subsubsection{Scenario of Core Growth over Time}
\label{subsubsec:growth_scenario}

The first scenario aligns with the dynamical mass accretion paradigm, which posits that young stellar objects grow primarily through prolonged, dynamical mass accretion from their natal, collapsing clump-scale reservoirs, rather than via isolated monolithic core collapse. A central prediction of this framework is that more evolved protostellar systems are systematically more massive, having benefited from longer accretion histories. This idea is foundational to the competitive accretion model \citep{2001MNRAS.323..785B,2003MNRAS.343..413B}, wherein protostars embedded in a common, turbulent, gravitationally collapsing gas reservoir accrete at rates proportional to their instantaneous mass, leading to accelerated growth for centrally located objects. Our results—showing enhanced mass and density of starless cores in IRBCs (late stage) relative to IRDCs (early stage)—are consistent with this picture. Additional observational support comes from studies of dense cores in IRDCs versus IRBCs with hub-filament structures, which report systematic increases in core mass and surface density interpreted as the result of hierarchical, multi-scale gas inflow \citep{2010ApJ...725...17G,2022MNRAS.510.5009L,2022MNRAS.511.3618L,2023MNRAS.522.3719L,2023MNRAS.520.3259X,2023ApJ...953...40Y}. Furthermore, recent statistical analyzes of dense cores in diverse regions—including the Dragon IRDC \citep{2021ApJ...912..156K}, the Orion Molecular Cloud \citep{2023ApJS..264...35T}, the ASHES IRDC sample \citep{2023ApJ...949..109L}, the ALMA-IMF survey \citep{2023A&A...674A..75N,2023A&A...674A..76P}, the ASSEMBLE survey \citep{2024ApJS..270....9X}, and the ALMAGAL survey \citep{2025A&A...696A.151C}—consistently find that protostellar cores are significantly more massive than starless cores, strongly suggesting mass growth over time.

Regarding timescales, \citet{2008ApJ...684.1240E} derived a mean lifetime of \(0.5 \pm 0.3\) Myr for starless/prestellar cores (\(0.2\)–\(10\,M_{\odot}\)) in low-mass star-forming regions, while a similar lifetime has been found for prestellar cores in high-mass regions like the ASHES sample \citep{2026ApJ...997..155M}. These lifetimes are several times the free-fall time of dense cores. Following \citet{2025A&A...696A..11V}, the lifetime of QUARKS starless cores is estimated to be \(\sim (1\)–\(8) \times 10^{5}\) yr. Moreover, \citet{2021MNRAS.505.2801L} determined that at least 47\% of the protocluster clumps targeted by QUARKS are already in the compact H\,\textsc{ii} region phase, which has a characteristic lifetime of \(\sim 10^{5}\) yr \citep{1989ApJS...69..831W,2011MNRAS.416..972D}. Taken together, these constraints suggest that the lifetimes of starless cores (primarily below \(10\,M_{\odot}\)) in both IRDC and IRBC environments investigated here are comparable to or shorter than the evolutionary timescales of their natal clumps. We therefore suggest that starless cores in both ASHES and QUARKS samples have sufficient time to grow in mass and density through dynamical mass accretion as their host clumps evolve from the early IRDC to the late IRBC stage.
For instance, at a gas infall rate of $\rm\sim10^{-4}\,M_{\odot}\,yr^{-1}$ typical of high-mass star formation (e.g., \citealt{2018ApJ...852...12Y,2022MNRAS.510.5009L,2025ARA&A..63....1B,2026arXiv260201238G}
), a $1\,M_{\odot}$ core would gain approximately $10\,M_{\odot}$ over a timescale of $\sim10^5$\,yr.

\subsubsection{Scenario of core formation under distinct initial conditions}
\label{subsubsec:growth_scenario2}
From the ALMAGAL survey’s comprehensive analysis of fragmentation properties across 1,013 high-mass star-forming clumps, \cite{2025A&A...696A.151C} found not only that core mass increases with evolutionary stage, but also that a population of lower-mass cores persists at all stages. This implies that dense cores can form continuously throughout a clump’s evolution—not solely during its earliest phases. Inspired by this, we propose that some QUARKS starless cores may have formed recently within the more evolved, turbulence-enhanced IRBC environments, rather than being direct descendants of earlier-stage IRDC cores. This proposal is supported by two complementary lines of evidence.

First, the dynamical state of QUARKS starless cores reflects their birth environment. As shown in Figure\,\ref{fig:alpha_tot:m}, QUARKS cores exhibit a significantly higher median total virial parameter (\(\alpha_{\rm vir,tot} \sim 2.3\)) compared to ASHES cores (\(\sim 1.0\)). This suggests that QUARKS starless cores are born under conditions of elevated turbulence, rather than inheriting a quiescent state from earlier epochs. In addition, the higher Jeans mass in IRBC environments—due to elevated temperatures and turbulence—naturally favors the formation of more massive cores. Thus, the observed masses of QUARKS starless cores could not arise solely from prolonged accretion; they could also reflect a shift in the initial fragmentation scale driven by the evolving physical conditions of the natal clump.

 \begin{figure}[ht!]
    \centering
    \includegraphics[angle=0, width=0.49\textwidth]{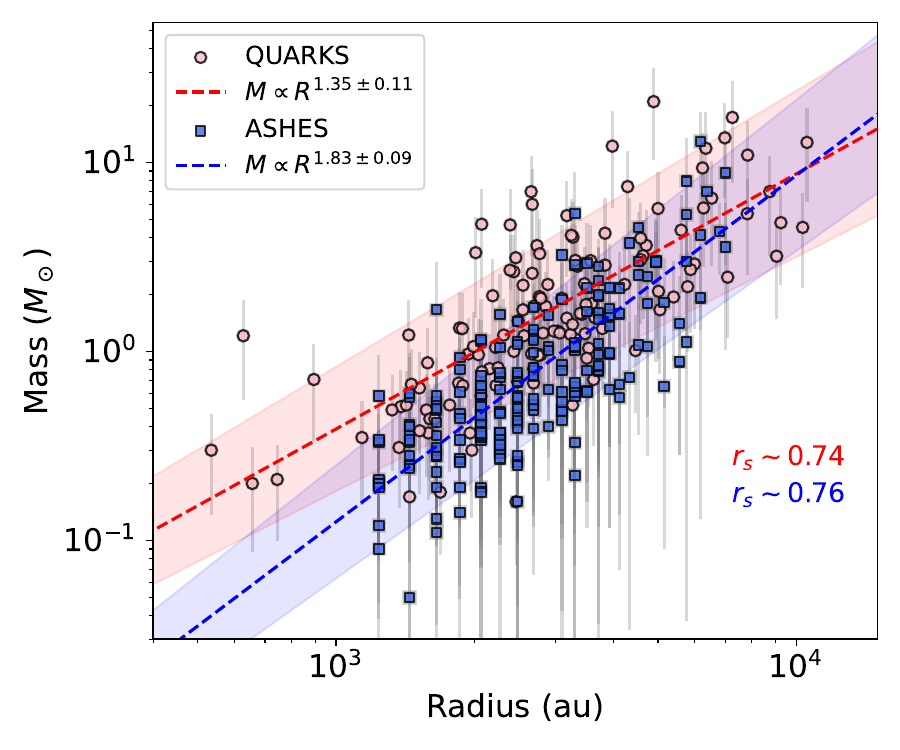}  
    \caption{Mass versus radius ($M-R$) diagram. The core masses for the ASHES survey (filled squares) were derived from the $T_{\rm NH_3}$, and for the QUARKS survey (filled circles) estimated from a fixed 20\,K dust temperature assumption. The error bars indicate the standard deviation from 1000 Monte Carlo simulations of each core (see Appendix\,\ref{app:b}).
    } 
    \label{fig:mr}    
\end{figure}

Second, the structural properties of these cores differ systematically. Figure~\ref{fig:mr} shows the Mass ($M$)–Radius ($R$) relation for both samples. A linear fit
yields \(M \propto R^{1.83 \pm 0.09}\) for ASHES cores (correlation coefficient \(r \sim 0.76\)), corresponding to an average density profile of \(\rho \propto R^{-1.2}\) (assuming spherical symmetry and uniform density). In contrast, QUARKS cores follow \(M \propto R^{1.35 \pm 0.11}\) (\(r \sim 0.74\)), implying a steeper density profile of \(\rho \propto R^{-1.7}\). Although the absolute slope values are sensitive to measurement uncertainties, 
the difference between the two slopes is statistically significant (\(>3\sigma\), \(\sigma \sim 0.1\)), supporting a transition toward steeper density gradients of starless cores in evolved IRBC clumps. Similarly, such result can also be found in \citealt{2023MNRAS.522.3719L}, where  comparative studies of hub-filament systems further show that IRBC hubs exhibit more compact, spherical dust emission than their IRDC counterparts (see Figure.\,1 and A1 of \citet{2023MNRAS.522.3719L}). Together, these observations suggest that the initial conditions for core formation differ between IRDCs and IRBCs: in cold, low-turbulence IRDCs (ASHES), cores form with lower initial masses, lower central densities, and shallower density profiles; whereas in warm, turbulence-enhanced IRBCs (QUARKS), cores form with higher initial masses, higher central densities, and steeper radial profiles.

\subsection{Implication for a dynamical mass accretion scenario of high-mass star formation}
The observed mass distribution of starless cores in both early- and late-stage high-mass star-forming regions provides a crucial empirical test for competing theoretical models. Currently, the turbulent core accretion \citep{2003ApJ...585..850M} and competitive accretion \citep{2001MNRAS.323..785B} scenarios represent the two leading paradigms. The former posits that high-mass stars form via monolithic collapse of massive prestellar (i.e., gravitationally bound starless) cores with initial masses above a few $ 10\times \,M_\odot$ \citep{2013ApJ...779...96T,2014prpl.conf..149T}, whereas the latter proposes that clumps first fragment into numerous low-mass, thermal-Jeans-scale cores ($\sim 1\,M_\odot$), which subsequently grow to high masses through prolonged, competitive accretion from their extended natal reservoirs—probably mediated by filamentary networks \citep{2019MNRAS.490.3061V,2020ApJ...900...82P}.

Over the past decade, several candidate high-mass prestellar cores (typically $M \gtrsim 16\,M_\odot$) have been reported \citep{2014MNRAS.439.3275W,2023A&A...675A..53B,2024ApJ...961L..35M,2025A&A...696A..11V,2025ApJS..280...33Y,2025ApJ...995..193Y}. However, none has been unambiguously confirmed as truly prestellar and massive at birth. Critically, most candidates reside in evolved clumps, making it impossible to distinguish whether they formed massive initially or grew significantly via accretion, which is a key degeneracy highlighted in Sect.\,\ref{subsubsec:growth_scenario}. Combing both ATOMS 3\,mm and QUARKS 1.3\,mm dust continuum data \citep{2025ApJ...995..193Y}, we have tentatively identified a candidate high-mass prestellar core of $27\,M_\odot$ within a radius of ~2800au, which is found as an early-phase object due to its location within an IR-dark lane within the IRAS 18290-0924 clump. This core remains a rare potential candidate supporting the turbulent core model.


Strikingly, the vast majority of starless cores are low-mass, regardless of the evolutionary stage of their host clumps.
Specifically, QUARKS starless cores span $0.2$–$21\,M_\odot$ with a median of $1.5\,M_\odot$, while ASHES cores range from $\sim 0.1$ to $12\,M_\odot$ with a median of $0.6\,M_\odot$. Crucially, almost none (see below for the exception in the QUARKS sample) of these cores from both samples  possess sufficient mass ($\gtrsim 16\,M_\odot$) to form a high-mass star via monolithic collapse, even when accounting for core-to-star efficiencies of $\sim 30$–$50\%$. 
Particularly, we identify from the QUARKS sample two high-mass starless cores with masses of  $17.3\,M_\odot$ and $21\,M_\odot$, which are above the $16\,M_\odot$ threshold for high-mass star formation. Although both cores are not among the 27 potentially misclassified protostellar candidates (see Sect.\,\ref{sec:limit}), they lie within the UCHii clumps; and thus their high-mass origins require further confirmation to distinguish between the core grow over time (see Sect.\,\ref{subsubsec:growth_scenario}) and  new formation in evolved natal clumps (see Sect.\,\ref{subsubsec:growth_scenario2}).


The systematic lack of massive starless cores in our sample suggests that such cores are initially rare in clustered environments, and thus pure ``turbulent core accretion" could not be the dominant pathway for high-mass star formation therein.
Instead, it favors models in which high-mass stars assemble their mass dynamically over time. Indeed, abundant evidence shows that gas flows along filaments and streamers can efficiently feed central cores \citep{2022MNRAS.516.1983S,2022MNRAS.514.6038Z,2023ApJ...953...40Y,2023MNRAS.520.3259X,2023MNRAS.522.3719L,2024ApJ...960...76P,2025A&A...700A..47Z,2025A&A...704A.149G,2025arXiv250915527M}, enabling sustained accretion rates compatible with competitive-like, dynamical mass accretion scenarios \citep{2018ARA&A..56...41M}. Our findings thus support a picture wherein high-mass star-forming clumps first fragment into predominantly low-mass cores, and high-mass stars emerge primarily through a dynamical mass accretion from beyond their initial core boundaries—with monolithic collapse from a massive prestellar core being, at best, a rare alternative.


\subsection{Limitations}\label{sec:limit}

From systematic comparative analysis for starless cores using ASHES and QUARKS surveys, we have concluded that starless cores are on average more massive ( by a factor of 2.4) and denser (by a factor of 1.7) in IRBC environments than in IRDC environments. However, several limitations should be noted.

First, different core extraction methods (\texttt{astrodendro} for ASHES vs. \texttt{getsf} for QUARKS) may introduce systematic biases to our comparative analysis. We therefore investigated in Appendix\,\ref{app:c} the potential impact of different algorithms on related analysis. As a result, we found that the significant differences in mass and density distributions between the starless cores from the ASHES and QUARKS surveys are not an artificial result from different core identification methods.

It is worth noting that in this work we prefer to adopt the \texttt{astrodendro} cores (i.e., 197 sources) instead of  \texttt{getsf} ones (3.5 times less, see Appendix\,\ref{app:c}) for the ASHES survey since the former population can provide sufficient ASHES sample for systematic comparative analysis with the QUARKS sample. In addition, we opted to use the \textit{getsf} cores for the QUARKS survey to maintain consistency with the core catalogs published and in preparation by the QUARKS survey team (\citealt{2025ApJS..280...33Y}; W. Y. Jiao et al. 2026, in prep.). This choice ensures alignment with ongoing survey-wide analyses, even at the cost of introducing a minor methodological heterogeneity in the cross-survey comparison. Although we have carefully addressed the potential biases introduced by the use of different core extraction methods, a fundamental methodological limitation remains. That is, conducting a direct, parameter-by-parameter comparison between cores identified with two distinct algorithms inherently compromises full consistency. A more robust approach for future studies would be to apply a single, uniform extraction tool—such as \texttt{astrodendro}, which was used for the ASHES data—to the QUARKS dataset as well, thereby enabling a truly standardized comparative analysis.

Second, the classification of starless cores may be influenced by differences in spectral line sensitivity between the ASHES and QUARKS surveys. While both surveys achieved comparable sensitivities ($\sim 10$\,mJy~beam$^{-1}$, see Table\,\ref{tab:almaspw}) for outflow tracers investigated here (i.e., CO and SiO), the ASHES data are approximately three times more sensitive than QUARKS for warm gas tracers (i.e., H$_2$CO and CH$_3$OH). Consequently, some very low-luminosity protostars—potentially driving only weak outflows or emitting faint warm molecular lines—may have been misclassified as starless cores in the QUARKS sample. Since protostellar cores are typically more massive than true starless cores, such contamination could systematically elevate the high-mass end of the QUARKS core mass function. To assess this potential bias, we re-examined our QUARKS starless core sample using higher-resolution ($\sim 0.3\arcsec$) and more sensitive ($\sim3$\,mJy~beam$^{-1}$) ALMA TM1 continuum and line data from the same QUARKS survey (\citealt{2024RAA....24b5009L}; W. Y. Jiao et al., in prep.). This deeper inspection revealed that 27 out of 127 (21.3\%) of the originally classified starless cores show signatures of either weak outflows or warm molecular gas emission, indicating they are likely protostellar in nature. We therefore repeated our comparative analysis by examining mass and density distributions, the CMF and $M-R$ relations using a refined QUARKS sample that excludes these 27 reclassified sources. The trends obtained with the full sample retain in the refined sample.
For instance, the refined QUARKS subsample shows median masses and densities decreased slightly by $\sim6\%$ and $15\%$, respectively, compared with those derived from the full sample. 
A one-sided Kolmogorov–Smirnov test confirms that the differences in both mass and density distributions remain highly significant ($p \ll 0.05$). 
Additionally, for the refined (‘clean’) QUARKS cores, their M-R slope steepens from $1.35\pm0.11$ to $1.47\pm0.13$, but its relation relative to that for the ASHES cores still holds (see Sect.\ref{subsubsec:growth_scenario2}).

Third,  the impact of dust temperature assumptions on the derived mass distributions of the two
starless core sample may not be ignored in analysis. Such impact has been partially taken into account in Sect.\,\ref{sec:3.2}. Note that for all QUARKS starless cores in IRBC clumps, we adopted therein a fixed  $T_{\rm dust} = 20\,\mathrm{K}$ to compute temperature-dependent quantities such as mass and density. However, IRBCs often host luminous protostars and H\,\textsc{ii} regions that would induce temperature gradients across the clump, raising concerns that a single temperature may introduce systematic uncertainties. To assess whether a realistic temperature gradient—approximately spanning 10–30 K across an IRBC clump—could substantially alter the derived trends in core properties, we performed MC simulations in which the dust temperature for each core was drawn uniformly from this range, rather than being fixed at 20 K with a narrow distribution around it. After such attempts, we find that the observed difference in core mass distributions
between the two samples remains robust and not an artifact of dust temperature assumptions (see Sect.\,Appendix\,\ref{app:b} for details).

In summary, even after accounting for potential protostellar contamination, potential biases from different core-extraction methods and temperatures assumptions, the systematic enhancement in core mass and density in IRBCs relative to IRDCs persists robustly, reinforcing our major conclusions. Forthcoming studies with both high-sensitivity and resolution observations are required to definitely characterize  the physical properties of true starless cores in different environments for a better understanding of the physical origin of starless cores in two markedly distinct star-forming environments.

\section{Summary and Conclusions}\label{sec:5}

Using the ALMA data at 1.3mm from both ASHES and QUARKS surveys, we have conducted a systematic comparative analysis of the physical properties—including radius, mass, density, kinematics, and dynamical state—of 324 candidate starless cores identified in early-phase infrared-dark clouds (IRDCs; ASHES survey, \citealt{2019ApJ...886..102S,2023ApJ...949..109L}) and evolved-phase infrared-bright clouds (IRBCs; QUARKS survey, \citealt{2025ApJS..280...33Y}). Our major findings are summarized as follows:

(1) Despite having comparable sizes (\(\sim 2500\) au), starless cores in IRBCs exhibit systematically higher masses (median \(1.5\,M_{\odot}\) vs. \(0.6\,M_{\odot}\)), number densities (\(2.0 \times 10^{6}\,\mathrm{cm}^{-3}\) vs. \(1.2 \times 10^{6}\,\mathrm{cm}^{-3}\)), and surface densities (\(0.5\,\mathrm{g\,cm}^{-2}\) vs. \(0.3\,\mathrm{g\,cm}^{-2}\)) than those in IRDCs. These enhancements, approximately a factor of two across all three parameters, are robustly confirmed by the cumulative core mass function, which shows the QUARKS population consistently above the ASHES population.

(2) Molecular line observations (H\(^{13}\)CO\(^+\), N\(_2\)D\(^+\), DCN, H\(_2\)CO) reveal that starless cores in IRBCs possess significantly stronger turbulent motions (\(\sigma_{\mathrm{obs, int}} \sim 0.5\,\mathrm{km\,s}^{-1}\)) compared to those cores in IRDCs (\(\sim 0.3\,\mathrm{km\,s}^{-1}\)). Consequently, starless cores in IRBCs display a higher median total virial parameter (\(\alpha_{\mathrm{vir,tot}} \sim 2.3\)) than those cores in IRDCs (\(\sim 1.0\)), with only \(42.3\%\) of QUARKS cores satisfying \(\alpha_{\mathrm{vir,tot}} < 2\) (a common threshold for gravitational binding), compared to \(75.3\%\) in ASHES. The elevated \(\alpha_{\mathrm{vir,tot}}\) of starless cores in IRBCs is attributed to enhanced turbulence likely driven by feedback from ongoing star formation (e.g., outflows, H\,\textsc{ii} regions), as further evidenced by the large intrinsic scatter in the \(\alpha_{\mathrm{vir,tot}}\)–mass relation.

(3) In the mass–radius (\(M\)–\(R\)) diagram, both samples follow power-law scaling relations, but the QUARKS cores exhibit a significantly flatter slope (\(M \propto R^{1.35 \pm 0.11}\)) than ASHES cores (\(M \propto R^{1.83 \pm 0.09}\)), with the difference exceeding \(3\sigma\). This implies steeper average density profiles for starless cores in IRBCs (\(\rho \propto R^{-1.7}\)) compared to starless cores in IRDCs (\(\rho \propto R^{-1.2}\)), consistent with more centrally concentrated structures in evolved environments.

In conclusion, these results support a dual evolutionary picture: (i) dense cores can form anew in evolved, turbulent IRBC environments, not solely during the earliest IRDC phase; and (ii) the initial conditions in IRBCs—characterized by warmer temperatures, stronger turbulence, and steeper density gradients—shift the fragmentation scale toward higher masses and densities relative to IRDCs. Furthermore, the observed prevalence of low-mass starless cores within both early IRDC and late IRBC environments challenges the necessity of massive initial cores and instead underscores the importance of environment-mediated, time-dependent mass growth through dynamical mass accretion from extended reservoirs (e.g., filaments and clump-scale flows), with monolithic collapse from a massive core being at most a rare occurrence. Thus, both new formation under altered initial conditions and subsequent dynamical growth are essential to explain the observed properties of starless cores across the evolutionary sequence of high-mass star formation.




\section*{Acknowledgements}
This work has been supported by the National Key R\&D Program of China (No.\,2022YFA1603101), the National Natural Science Foundation of China (NSFC) through grants No.12073061, No.12122307, and No.12033005.
H.-L. Liu is supported by Yunnan Fundamental Research Project (grant No.\,202301AT070118, 202401AS070121), and by Xingdian Talent Support Plan--Youth Project. 
Tie Liu acknowledges the supports by the PIFI program of Chinese Academy of Sciences through grant No. 2025PG0009, and the Tianchi Talent Program of Xinjiang Uygur Autonomous Region.
PS was partially supported by a Grant-in-Aid for Scientific Research (KAKENHI Number JP23H01221) of JSPS.
LB gratefully acknowledges support by the ANID BASAL project FB210003.
SRD acknowledges support from the Fondecyt Postdoctoral fellowship (project code 3220162) and ANID BASAL project FB210003.
G.G. gratefully acknowledges support by the ANID BASAL project FB210003.
AS gratefully acknowledges support by the Fondecyt Regular (project
code 1220610), ANID BASAL project FB210003, and the China-Chile Joint Research
Fund (CCJRF No. 2312).
S.L. acknowledges support from the National SKA Program of China with No. 2025SKA0140100, “Double First-Class” Funding with No. 14912217, and National Natural Science Foundation of China (NSFC) grant with No. 13004007.
C.W.L is supported by the Basic Science Research Program through the NRF funded by the Ministry of Education, Science and Technology (grant No. NRF-2019R1A2C1010851) and by the Korea Astronomy and Space Science Institute grant funded by the Korea government (MSIT; project No. 2025-1-841-02).
This paper makes use of the following ALMA data: ADS/JAO.ALMA\#2021.1.00095.S.

\appendix
\section{Impact of Different Core Identification Methods}\label{app:c}

\begin{figure*}[ht!]
    \centering
    \includegraphics[angle=0, width=1\textwidth]{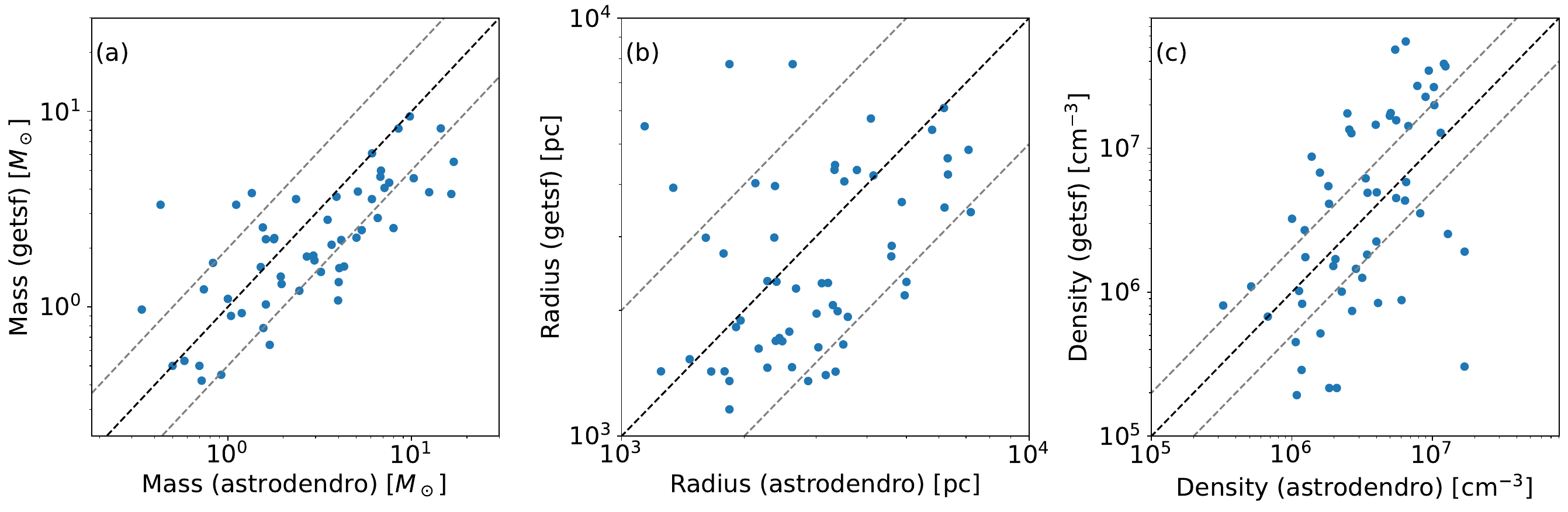}  
    \caption{Masses, radii and number densities of the 56 ASHES cores derived from the \texttt{astrodendro} and \texttt{getsf}. The black dashed line represents $ y = x$, while the two gray dashed lines denote $y = 0.5x$ and $y = 2x$, respectively.} 
    \label{fig:astr_getsf}    
\end{figure*}

We evaluated the potential impact of different core identification algorithms (\texttt{astrodendro} and \texttt{getsf}) on related analysis of the ASHES and QUARKS sample. To this end, we primarily accessed the impact of different identification algorithms on the ASHES sample. 
We first cross-matched the 66 dust continuum cores previously identified by \cite{2024ApJS..270....9X} using \texttt{getsf} from the ASHES pilot survey with the 294 cores identified by \cite{2019ApJ...886..102S} using \texttt{astrodendro} from the same survey, resulting in 56 common sources.
Subsequently, we computed following  \cite{2019ApJ...886..102S} the masses, radii, and number densities of these 56 cores.  As shown in Fig.\,\ref{fig:astr_getsf}, the masses and radii of \texttt{astrodendro} cores are overall significantly higher—by roughly a factor of two—than those \texttt{getsf} counterparts. This difference is reinforced by the one-sided K-S tests with $p$-values of about 0.03 for the distributions of both parameters. Also, this finding is consistent with recent results from \cite{2026ApJ...997..155M}. 
Moreover, the one-sided K-S test of the number density produced a $p$-value of $\sim$0.12, indicating no significant difference between the \texttt{astrodendro} and  \texttt{getsf} cores.

\begin{figure*}[ht!]
    \centering
    \includegraphics[angle=0, width=0.5\textwidth]{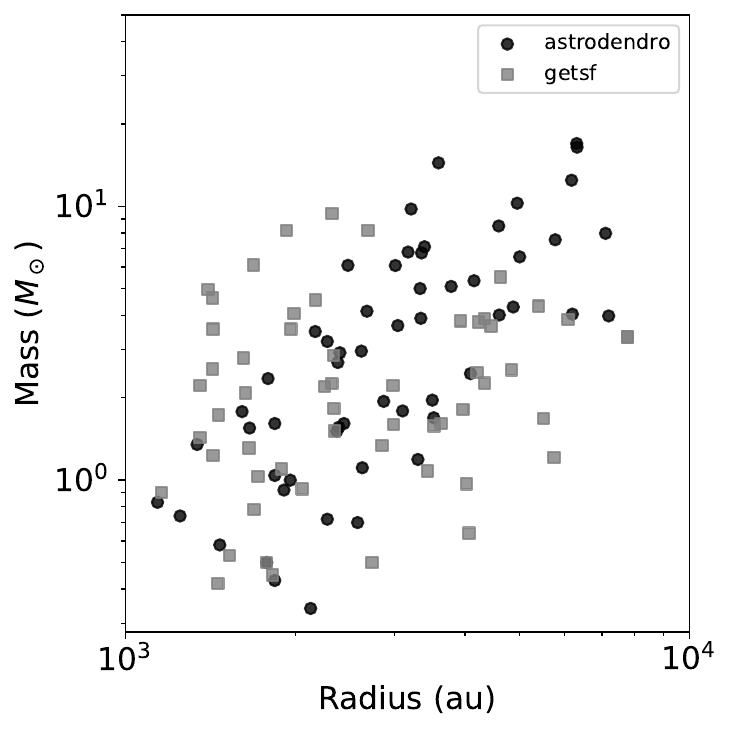}  
    \caption{Mass versus radius ($M-R$) diagram for 56 dense cores identified from \texttt{astrodendro} (black dots) and \texttt{getsf} (gray squares), respectively.} 
    \label{fig:astr_getsf_m_r}    
\end{figure*}

Furthermore, we present in Fig.\,\ref{fig:astr_getsf_m_r} the $M-R$ relations for the 56 common ASHES cores identified using the two different methods. Both \texttt{astrodendro} and  \texttt{getsf} cores appear to follow similar $M-R$ trends. However,
the one-side K-S test for the two populations of cores cannot quantitatively confirm their similarity in the M-R relations, which  
could be due to the limited dynamic range in mass for the  \texttt{getsf} cores (i.e., no sample points for \texttt{getsf} sources in the right upper corner of the figure). 


Taken together, we therefore suggest that the differences in mass and density distributions between the starless cores from  the QUARKS and ASHES surveys are not attributable to the core identification methods. Particularly,  correcting for the influence of the core extraction algorithm will widen the overall gap in mass distributions between the ASHES and QUARKS starless cores, thereby facilitating comparative analysis of the core mass function  between ASHES and QUARKS samples (see Sect.\,\ref{subsec:3.5}). While we cannot entirely rule out a subtle influence of methodological differences on the derived mass–radius ($M$–$R$) relations, the seemingly similar $M$–$R$ trends seen in the two core populations (see Fig.\,9)—despite being identified with different algorithms—are consistent with their nearly identical density distributions as mentioned earlier. This further supports the robustness of our comparative analysis and reinforces the physical origin of the observed evolutionary trends.

\section{Effect of  Distance and Sample Number on Statistical Significance}\label{app:a}
We examined here whether the observed differences in physical parameters (radius and mass) of starless cores between the ASHES and QUARKS samples arise from their differing clump distances and sample number or not?

\begin{figure*}[ht!]
    \centering
    \includegraphics[angle=0, width=1\textwidth]{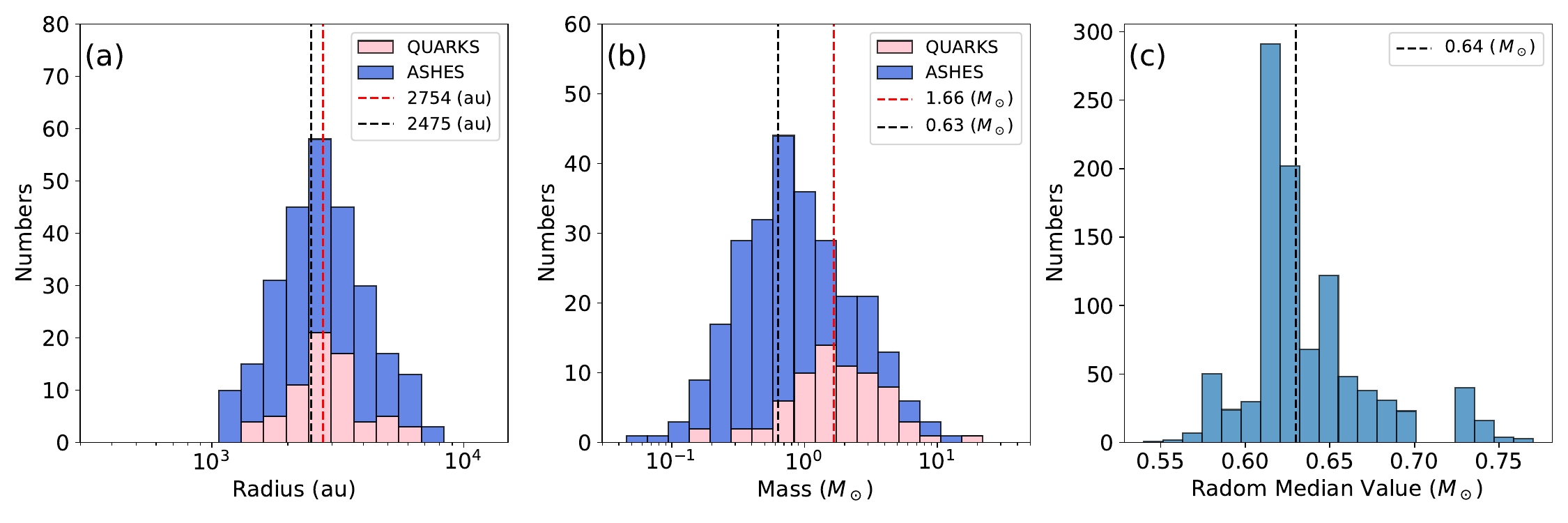}  
    \caption{Distributions of the radius (panel\,a) and mass (panel\,b) of the starless cores in the subset of the QUARKS sample and the ASHES sample, both of which are imposed to the same distance range.  Distribution of median core mass obtained from 1000 random resamplings for ASHES starless cores (panel\,c). The dashed lines in all panels indicate the median values for the corresponding distributions.} 
    \label{fig:prodis}    
\end{figure*}

ASHES clumps are located at approximately 2.9–5.4\,kpc, while the QUARKS sample spans a wider range of 1.2–11.6\,kpc. To evaluate bias due to distance variation, we restricted the QUARKS starless core sample to the same distance range as ASHES (2.9–5.4\,kpc), which produces a subset of 70 cores. We  compared the radii of the starless cores in the subset of the QUARKS sample and the ASHES sample (Figure\,\ref{fig:prodis}a). A two-sided K‑S test yields a $p$‑value >0.05, suggesting that the radii are drawn from the same distribution—a result consistent with previous findings (see Sect.\,\ref{sec:3.2}). Similarly, We did the comparison for the mass parameter (Figure\,\ref{fig:prodis}b). The systematic differences in the mass distribution still hold, where a median mass ratio between the two samples is $\sim$2.6 (1.66\,\msun\,/0.63\,\msun\,). A one-sided K‑S test yields a $p$‑value $\ll$0.05, indicating that these QUARKS cores are also significantly more massive than those of ASHES. Therefore, we suggest that the disparity in mass between the ASHES and QUARKS samples is not driven by the clump distance.

In addition, to evaluate potential biases arising from the differing sample number between the QUARKS and ASHES samples, we performed a resampling analysis. Specifically, we repeatedly drew random subsamples of 127 (counts of QUARKS starless cores) cores from the full set of 197 ASHES starless cores, conducting 1000 iterations of this process. Across all iterations, the median mass of the ASHES starless cores fell within the range of $\sim$0.5 to 0.8\,\msun\,, with a median of $\sim$0.6\,\msun\, over all randomly generated subsamples, as shown in Figure\,\ref{fig:prodis}c. Importantly, each randomly selected subset of the ASHES sample yielded a median mass significantly lower than that of the QUARKS sample (i.e., 1.66\,\msun), with one-sided K-S tests consistently returning $p$-values$\ll$ 0.05. These results are consistent with our earlier findings (see Sect.\,\ref{sec:3.2}), indicating that the difference in sample sizes is not responsible for the observed disparity in the mass distribution between the ASHES and QUARKS starless cores.

\section{Effect of Dust Temperature} \label{app:b}
We investigated here the impact of dust temperature assumptions on the derived mass distributions of the two starless core samples. Using a Monte Carlo (MC) approach, we simulated the mass distributions for both the ASHES and QUARKS starless cores by sampling dust temperatures over the specific ranges: 7.0–22.8\,K for ASHES (based on NH$_3$ line-derived temperatures) and 10–30\,K for QUARKS (reflecting the assumed range around the fiducial value of 20\,K). Specifically, for each simulation, we assigned a random temperature uniformly drawn from the respective range to every core in the sample and recalculated its mass accordingly. A total of  1,000 independent MC realizations were run.

\begin{figure*}[ht!]
    \centering
    \includegraphics[angle=0, width=1\textwidth]{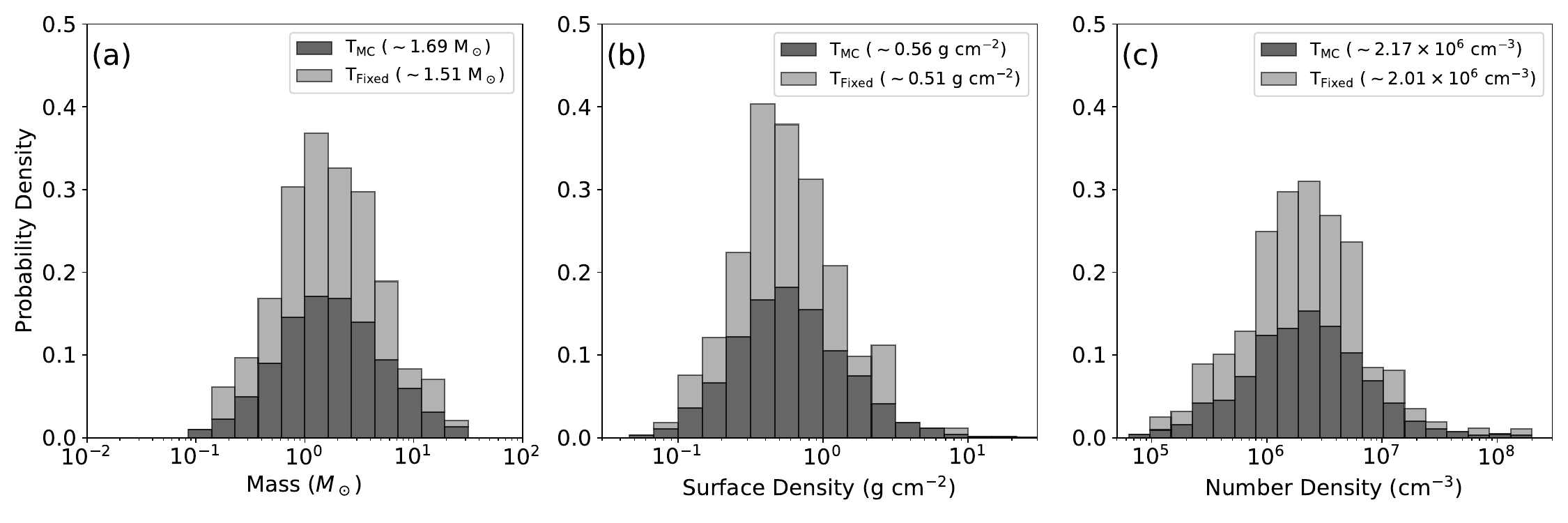}  
    \caption{Histograms of the relevant physical parameters for QUARKS starless cores, derived using a fixed dust temperature ($T_{\rm Fixed}$) of 20\,K (gray) and from Monte Carlo simulations where the dust temperature ($T_{\rm MC}$) was randomly sampled from a range of $10-30$\,K (black), respectively. Median values for each distribution are given in parentheses.
    } 
    \label{fig:pro_all}    
\end{figure*}

Note that we adopted a uniform sampling of dust temperatures in MC simulations, rather than sampling from a narrow distribution centered on a fixed value. This approach better accounts for potential temperature gradients within a clump—particularly relevant for IRBCs, which often harbor luminous protostars and H\,\textsc{ii} regions capable of inducing significant thermal structure across the clump. By uniformly sampling the full temperature range, we can evaluate whether assuming a single, representative dust temperature introduces systematic biases in the derived temperature-dependent parameters (mass, surface density, and number density) for QUARKS starless cores. As shown in Figure~\ref{fig:pro_all}, we directly compare the parameter distributions obtained using a fixed $T_{\rm dust} = 20\,\mathrm{K}$ versus those derived from randomly assigned temperatures drawn uniformly from the 10–30 K range. The resulting distributions are nearly indistinguishable, with highly consistent median values between two types of calculation for all three parameters. Quantitatively, two-sided Kolmogorov–Smirnov (K–S) tests yield $p$-values of $\sim$0.81 (mass), $\sim$0.18 (surface density), and $\sim$0.56 (number density), indicating no statistically significant difference between the two sets of distributions. We therefore suggest that the use of a single dust temperature for all QUARKS starless cores does not introduce appreciable systematic uncertainties in  analysis of temperature-dependent parameters in this study.

\begin{figure*}[ht!]
    \centering
    \includegraphics[angle=0, width=0.8\textwidth]{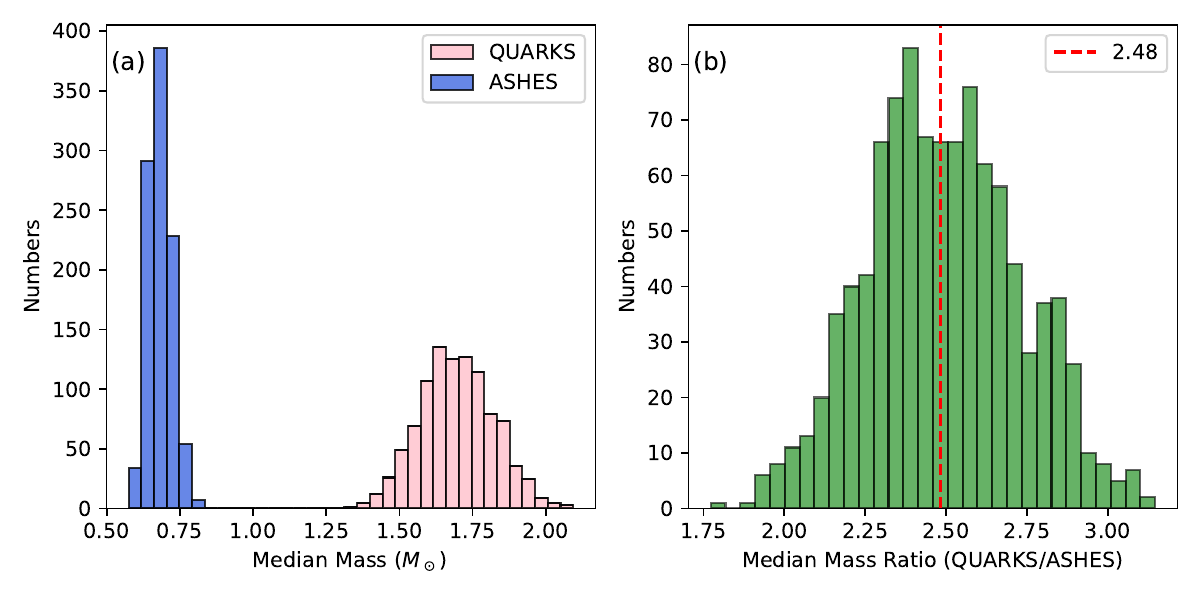}  
    \caption{Simulated mass distributions for both the ASHES and QUARKS starless cores (panel\,a) and the median mass ratio distribution between the simulated QUARKS and ASHES samples (panel\,b). The dashed line indicates the median value.
    } 
    \label{fig:simulation}    
\end{figure*}

Across all simulations, the median mass of QUARKS starless cores varies between $1.32\,M_\odot$ and $2.12\,M_\odot$, while that of ASHES cores ranges from $0.57\,M_\odot$ to $0.83\,M_\odot$ (Figure~\ref{fig:simulation}a). This yields a median mass ratio (QUARKS/ASHES) between $\sim 1.84$ and 3.42, with a median value of $2.48 \pm 0.25$ (Figure~\ref{fig:simulation}b). Furthermore, in every realization, a one-sided K–S test comparing the two mass distributions returns a $p$-value $\ll 0.05$ (not shown here), indicating that the QUARKS cores are consistently more massive than their ASHES counterparts at high statistical significance, regardless of the adopted temperature uncertainty. We therefore conclude that the observed disparity in core mass distributions between the two samples is robust and not an artifact of dust temperature assumptions.

\section{Parameters of molecular line emission from the QUARKS starless cores}
Table~\ref{table:line:pro} presents the relevant Gaussian‑fit parameters of $\rm H_2CO$, DCN, $\rm N_2D^+$, and $\rm H^{13}CO^+$ detected in the QUARKS starless cores.

\renewcommand{\thetable}{D\arabic{table}}   
\setcounter{table}{0}                       
\begin{deluxetable*}{cccccccccccccccccc}\label{table:line:pro}
\tabletypesize{\footnotesize}
\tablewidth{0pt} 
\tablecaption{Parameters of molecular line emission from the QUARKS starless cores} 
\tablehead{
\colhead{Source} & \colhead{Core} & \multicolumn{4}{c}{$\rm H_2CO$}   & \multicolumn{4}{c}{$\rm DCN$} & \multicolumn{4}{c}{$\rm N_2D^+$} &\multicolumn{4}{c}{$\rm H^{13}CO^+$} \\
\cline{3-18}
&  & Detc.&$T$ & $V_{LSR}$ & $\sigma_{\rm obs}$ &Detc. &$T$ & $V_{LSR}$ & $\sigma_{\rm obs}$ &Detc.& $T$ & $V_{LSR}$ & $\sigma_{\rm obs}$ & Detc.&$T$ & $V_{LSR}$ & $\sigma_{\rm obs}$  \\ 
&  && (K) & ($\rm km~s^{-1}$)  & ($\rm km~s^{-1}$) &&(K) & ($\rm km~s^{-1}$)  & ($\rm km~s^{-1}$) &&(K) & ($\rm km~s^{-1}$)  & ($\rm km~s^{-1}$) &&(K) & ($\rm km~s^{-1}$)  & ($\rm km~s^{-1}$) \\}
\colnumbers 
\startdata 
I08448-4343 & 9 & 1 & 4.8 & 2.22 & 0.79 & 1 & 0.68 & 3.16 & 0.59 & 1 & 1.57 & 2.8 & 0.9 & 2 & 1.76 & 2.73 & 0.87 \\
I08448-4343 & 20 & 0 & 0.0 & 0.0 & 0.0 & 0 & 0.0 & 0.0 & 0.0 & 0 & 0.0 & 0 & 0.0 & 0 & 0.0 & 0.0 & 0.0 \\
I09002-4732 & 8 & 1 & 4.1 & 6.32 & 0.54 & 1 & 5.39 & 6.52 & 0.81 & 0 & 0.0 & 0 & 0.0 & 0 & 0.0 & 0.0 & 0.0 \\
I09094-4803 & 6 & 1 & 1.19 & 76.93 & 0.8 & 0 & 0.0 & 0.0 & 0.0 & 0 & 0.0 & 0 & 0.0 & 1 & 1.8 & 76.64 & 0.86 \\
I09094-4803 & 7 & 1 & 1.14 & 76.87 & 0.71 & 0 & 0.0 & 0.0 & 0.0 & 0 & 0.0 & 0 & 0.0 & 0 & 0.0 & 0.0 & 0.0 \\
I09094-4803 & 9 & 1 & 1.23 & 75.13 & 1.32 & 0 & 0.0 & 0.0 & 0.0 & 0 & 0.0 & 0 & 0.0 & 1 & 0.91 & 75.55 & 0.97 \\
I12572-6316$\_$2 & 3 & 0 & 0.0 & 0.0 & 0.0 & 0 & 0.0 & 0.0 & 0.0 & 0 & 0.0 & 0 & 0.0 & 1 & 1.78 & 30.64 & 0.82 \\
I12572-6316$\_$2 & 4 & 1 & 1.53 & 28.32 & 0.9 & 1 & 0.26 & 28.5 & 0.81 & 1 & 0.25 & 29.1 & 0.89 & 1 & 0.78 & 29.51 & 0.99 \\
I13079-6218 & 5 & 1 & 1.7 & -41.19 & 1.56 & 1 & 0.49 & -40.8 & 0.73 & 0 & 0.0 & 0 & 0.0 & 1 & 0.92 & -41.24 & 0.4 \\
I13079-6218 & 6 & 1 & 0.72 & -39.8 & 0.86 & 1 & 0.62 & -40.6 & 0.62 & 0 & 0.0 & 0 & 0.0 & 1 & 2.26 & -40.4 & 0.8 \\
I13111-6228 & 11 & 0 & 0.0 & 0.0 & 0.0 & 0 & 0.0 & 0.0 & 0.0 & 0 & 0.0 & 0 & 0.0 & 1 & 2.16 & -38.9 & 0.8 \\
I13134-6242 & 3 & 0 & 0.0 & 0.0 & 0.0 & 1 & 0.61 & -29.6 & 0.57 & 0 & 0.0 & 0 & 0.0 & 2 & 1.53 & -30.19 & 0.67 \\
I13140-6226 & 12 & 0 & 0.0 & 0.0 & 0.0 & 0 & 0.0 & 0.0 & 0.0 & 1 & 0.71 & -34.51 & 0.63 & 1 & 1.67 & -34.49 & 0.95 \\
I13291-6229 & 10 & 0 & 0.0 & 0.0 & 0.0 & 0 & 0.0 & 0.0 & 0.0 & 0 & 0.0 & 0 & 0.0 & 0 & 0.0 & 0.0 & 0.0 \\
I13291-6229 & 12 & 0 & 0.0 & 0.0 & 0.0 & 0 & 0.0 & 0.0 & 0.0 & 0 & 0.0 & 0 & 0.0 & 0 & 0.0 & 0.0 & 0.0 \\
I13291-6229 & 13 & 0 & 0.0 & 0.0 & 0.0 & 0 & 0.0 & 0.0 & 0.0 & 0 & 0.0 & 0 & 0.0 & 0 & 0.0 & 0.0 & 0.0 \\
I13295-6152 & 7 & 1 & 0.52 & -46.1 & 1.3 & 0 & 0.0 & 0.0 & 0.0 & 0 & 0.0 & 0 & 0.0 & 1 & 2.5 & -44.77 & 0.96 \\
I13295-6152 & 10 & 1 & 0.95 & -44.16 & 0.91 & 0 & 0.0 & 0.0 & 0.0 & 0 & 0.0 & 0 & 0.0 & 1 & 2.81 & -44.47 & 0.83 \\
I14050-6056 & 7 & 1 & 0.74 & -48.7 & 0.83 & 0 & 0.0 & 0.0 & 0.0 & 0 & 0.0 & 0 & 0.0 & 0 & 0.0 & 0.0 & 0.0 \\
I14050-6056 & 9 & 0 & 0.0 & 0.0 & 0.0 & 0 & 0.0 & 0.0 & 0.0 & 0 & 0.0 & 0 & 0.0 & 0 & 0.0 & 0.0 & 0.0 \\
I14206-6151 & 2 & 0 & 0.0 & 0.0 & 0.0 & 0 & 0.0 & 0.0 & 0.0 & 0 & 0.0 & 0 & 0.0 & 0 & 0.0 & 0.0 & 0.0 \\
I14206-6151 & 5 & 0 & 0.0 & 0.0 & 0.0 & 0 & 0.0 & 0.0 & 0.0 & 1 & 1.16 & -48-91 & 0.78 & 1 & 3.22 & -49.0 & 0.38 \\
I14206-6151 & 6 & 1 & 1.22 & -50.97 & 0.62 & 0 & 0.0 & 0.0 & 0.0 & 0 & 0.0 & 0 & 0.0 & 2 & 2.39 & -48.89 & 0.48 \\
I14212-6131 & 5 & 1 & 0.72 & -50.51 & 0.55 & 0 & 0.0 & 0.0 & 0.0 & 1 & 0.91 & -51.05 & 0.68 & 1 & 3.4 & -50.19 & 1.14 \\
I14212-6131 & 11 & 0 & 0.0 & 0.0 & 0.0 & 0 & 0.0 & 0.0 & 0.0 & 0 & 0.0 & 0 & 0.0 & 1 & 2.45 & -50.85 & 0.8 \\
I15384-5348 & 6 & 0 & 0.0 & 0.0 & 0.0 & 0 & 0.0 & 0.0 & 0.0 & 0 & 0.0 & 0 & 0.0 & 0 & 0.0 & 0.0 & 0.0 \\
I15384-5348 & 8 & 0 & 0.0 & 0.0 & 0.0 & 0 & 0.0 & 0.0 & 0.0 & 0 & 0.0 & 0 & 0.0 & 0 & 0.0 & 0.0 & 0.0 \\
I15384-5348 & 9 & 1 & 0.59 & -43.0 & 0.87 & 0 & 0.0 & 0.0 & 0.0 & 0 & 0.0 & 0 & 0.0 & 0 & 0.0 & 0.0 & 0.0 \\
I15384-5348 & 10 & 1 & 0.6 & -42.8 & 0.78 & 0 & 0.0 & 0.0 & 0.0 & 0 & 0.0 & 0 & 0.0 & 0 & 0.0 & 0.0 & 0.0 \\
I15384-5348 & 15 & 0 & 0.0 & 0.0 & 0.0 & 0 & 0.0 & 0.0 & 0.0 & 0 & 0.0 & 0 & 0.0 & 0 & 0.0 & 0.0 & 0.0 \\
I15411-5352 & 15 & 0 & 0.0 & 0.0 & 0.0 & 0 & 0.0 & 0.0 & 0.0 & 0 & 0.0 & 0 & 0.0 & 0 & 0.0 & 0.0 & 0.0 \\
I15437-5343 & 2 & 1 & 0.45 & -82.7 & 1.9 & 1 & 0.38 & -81.6 & 0.9 & 1 & 0.38 & -82.3 & 0.6 & 1 & 0.92 & -82.34 & 0.38 \\
I15437-5343 & 5 & 1 & 0.99 & -84.86 & 0.6 & 1 & 0.39 & -84.7 & 0.54 & 0 & 0.0 & 0 & 0.0 & 2 & 1.73 & -83.23 & 0.51 \\
\enddata
\begin{flushleft}
(The complete table is available in machine-readable form.)\\
\end{flushleft}
\end{deluxetable*}



\bibliographystyle{aasjournal}
\bibliography{reference}{}

\end{document}